\documentclass[12pt, draftclsnofoot, onecolumn]{IEEEtran}

\usepackage{xspace,amsmath,amssymb,amsfonts,epsfig,subfigure,syntonly}
\usepackage{cite,bm,color,url,textcomp,empheq,boxedminipage}
\usepackage{algorithmicx,algorithm}
\usepackage{epstopdf,makecell}
\usepackage{empheq}
\usepackage{pifont}
\usepackage{stfloats}

\usepackage{graphicx,graphics}  
\usepackage{multirow,multicol}
\usepackage{psfrag}    
\usepackage{stfloats}
\usepackage{url}
\usepackage{lipsum}

\newtheorem{remark}{\underline{Remark}}[section]

\newcommand{\mv}[1]{\mbox{\boldmath{$ #1 $}}}

\long\def\symbolfootnote[#1]#2{\begingroup
\def\thefootnote{\fnsymbol{footnote}}
\footnote[#1]{#2}\endgroup}

\psfull

\allowdisplaybreaks[4]

\begin{document}
\IEEEspecialpapernotice{(Invited Paper)}
\title{UAV-Enabled Wireless Power Transfer: A Tutorial Overview}

\author{Lifeng~Xie,~\IEEEmembership{Student Member,~IEEE,}~Xiaowen~Cao,~\IEEEmembership{Student Member,~IEEE,}~Jie~Xu,~\IEEEmembership{Member,~IEEE,}~and~Rui~Zhang,~\IEEEmembership{Fellow,~IEEE}
\thanks{L. Xie and X. Cao are with the Future Network of Intelligence Institute (FNii), The Chinese University of Hong Kong (Shenzhen), Shenzhen 518172, China, and the School of Information Engineering, Guangdong University of Technology, Guangzhou 510006, China  (e-mail: lifengxie@mail2.gdut.edu.cn, caoxwen@outlook.com).}
\thanks{J. Xu is with the FNii and the School of Science and Engineering, The Chinese University of Hong Kong (Shenzhen), Shenzhen 518172, China (e-mail: xujie@cuhk.edu.cn). J. Xu is the corresponding author.}
\thanks{R. Zhang is with the Department of Electrical and Computer Engineering, National University of Singapore, Singapore (e-mail: elezhang@nus.edu.sg).} 
}

\maketitle
\begin{abstract}
Unmanned aerial vehicle (UAV)-enabled wireless power transfer (WPT) has recently emerged as a promising technique to provide sustainable energy supply for widely distributed low-power ground devices (GDs) in large-scale wireless networks. Compared with the energy transmitters (ETs) in conventional WPT systems which are deployed at fixed locations, UAV-mounted aerial ETs can fly flexibly in the three-dimensional (3D) space to charge nearby GDs more efficiently. This paper provides a tutorial overview on UAV-enabled WPT and its appealing applications, in particular focusing on how to exploit UAVs' controllable mobility via their 3D trajectory design to maximize the amounts of energy transferred to all GDs in a wireless network with fairness. First, we consider the single-UAV-enabled WPT scenario with one UAV wirelessly charging multiple GDs at known locations. To solve the energy maximization problem in this case, we present a general trajectory design framework consisting of three innovative approaches to optimize the UAV trajectory, which are multi-location hovering, successive-hover-and-fly, and time-quantization-based optimization, respectively. Next, we consider the multi-UAV-enabled WPT scenario where multiple UAVs cooperatively charge many GDs in a large area. Building upon the single-UAV trajectory design, we propose two efficient schemes to jointly optimize multiple UAVs' trajectories, based on the principles of UAV swarming and GD clustering, respectively. Furthermore, we consider two important extensions of UAV-enabled WPT, namely UAV-enabled wireless powered communication networks (WPCN) and UAV-enabled wireless powered mobile edge computing (MEC), by integrating the emerging WPCN and MEC techniques, respectively. For both cases, we investigate the UAV trajectory design jointly with communication/computation resource allocations to optimize the system performance, subject to the energy availability constraints at GDs. Finally, open problems in UAV-enabled WPT and promising directions for its future research are discussed.
\end{abstract}

\section{Introduction}


The advancements in Internet of Things (IoT) are expected to enable numerous new applications in a variety of vertical domains such as smart city, smart factory, intelligent transportation systems, and so on. Towards this end, future wireless networks need to incorporate a massive number of low-power IoT devices with real-time sensing, communication, computation, and control functionalities. In this regard, how to maintain the sustainable operation of these low-power devices is becoming a more practically important as well as challenging problem to tackle. Different from conventional energy sources such as battery and/or environment energy harvesting\cite{optimalCKHo,energyoutage}, the radio-frequency (RF) transmission enabled wireless power transfer (WPT) has recently emerged as a viable new solution to provide sustainable energy supply for low-power IoT devices, where dedicated energy transmitters (ETs) are deployed for broadcasting RF signals to wirelessly charge them simultaneously \cite{SBi_WPTSurvey,Zhang19TIE_WPT,JXu16TSP_WPT,Huang14TWC_WPT,XLu15WC_WPT,Xu14TSP_onebit}.
Recently, there have been various start-up companies (such as Powercast, TransferFi, and Energous) that have developed commercialized RF-based WPT products for moderate-to-long-range wireless charging applications.
Furthermore, WPT has been integrated in wireless communication and computation \cite{Mao_MEC_Survery,Mach_MEC_Survery,XSun16Mag_EdgeIoT} networks for various new applications, such as simultaneous wireless information and power transfer (SWIPT) \cite{XZhou13TCOM_WIPT,RZhang13TWC_SWIPT,BClerckxJSAC19_EH,DNg13TWC_WIPT,JXuMISO}, wireless powered communication networks (WPCN) \cite{Lee16TWC_WPCN,SBi16WC_WPCN,HJu14TWC_WPCN}, and wireless powered mobile edge computing (MEC) \cite{Wang18TWC_MEC_WPT,Wang20TWC_WPMEC,SBi18TWC_WPMEC,XHu18TWC_WPMEC}.


How to enhance the energy transfer efficiency from ETs to distributed wireless devices is the essential challenge faced in RF-based WPT systems. First, due to the severe RF signal propagation loss over distance, the energy transfer efficiency degrades drastically when devices are located far away from the ETs. Second, when there are multiple devices distributed at different locations, the nearby devices from an ET will harvest significantly more energy than those far apart from all ETs, thus resulting in a critical near-far fairness issue. Under WPCN and wireless powered MEC applications, the near-far issue may become even more severe, in which the far-apart devices with less harvested energy need to consume more energy for information transmission to meet the same quality of service (QoS) as nearby devices.
To tackle these challenges, there have been prior works that proposed various techniques to enhance the energy transfer efficiency via, e.g., transmit energy beamforming \cite{LLiu14TSP_WPT,YZengTCOM17_WPT,DNg14TWC_Beam,Xu14TSP_onebit,JXu16TSP_WPT,GMa2021}, energy waveform optimization \cite{BClerckx16TSP_waveform,Clerckx18TSP_wave}, adaptive power control \cite{EBoshkovska17TCOM_WPT}, and deployment optimization \cite{SBi16TWC_WPT} for the ETs. Despite these research progresses, the ultra-dense deployment of ETs at fixed locations is in general necessary to achieve ubiquitous coverage for WPT. This thus incurs unduly high deployment and maintenance costs that hinder its broad applications in practice.

\begin{figure}
  \centering
  \includegraphics[width=10cm]{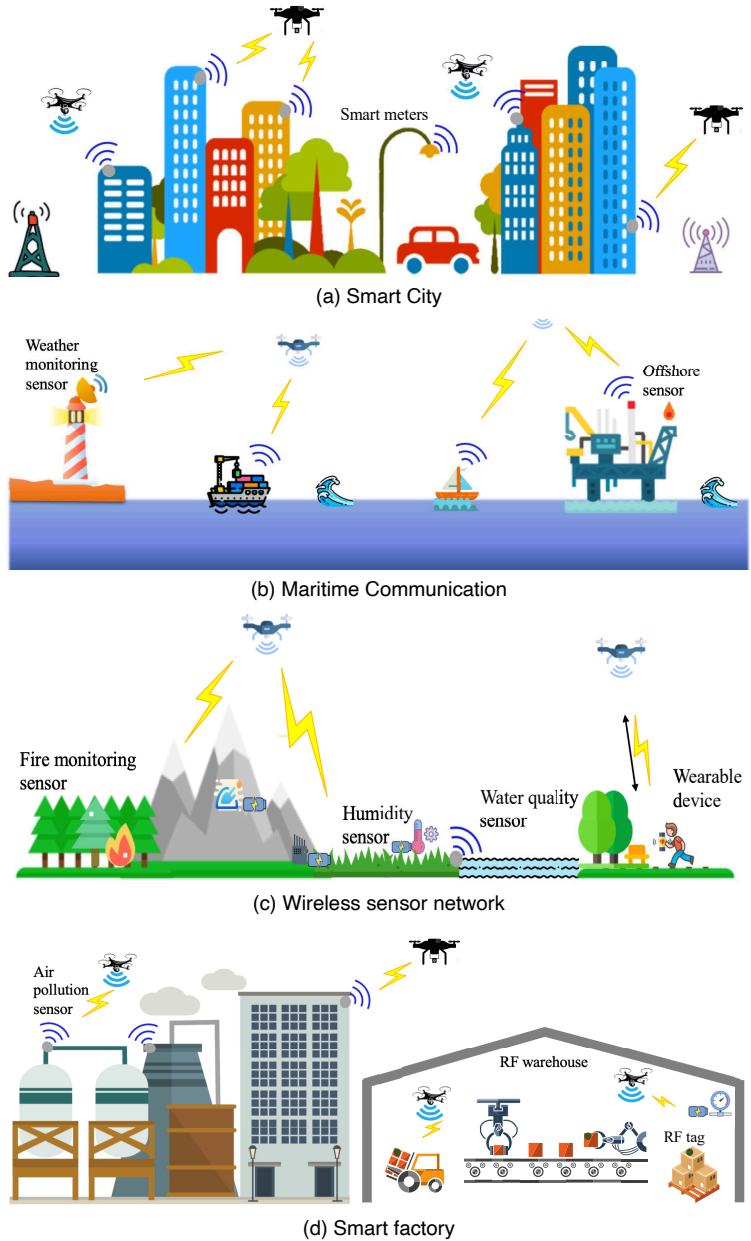}\\
  \caption{Example applications of UAV-enabled WPT.}\label{fig1}
  \vspace{-0.3cm}
\end{figure}


Recently, unmanned aerial vehicle (UAV)-enabled wireless communications have attracted growing interests (see, e.g.,\cite{YZengComM16_Survey,YZengWC19_CCUAV,Wu2020aa_JSAC20}), in which UAVs are dispatched as aerial base stations (BSs), access points (APs), relays, etc., to support the mobile subscribers on the ground for point-to-point communications \cite{ZengTCom16_Relay_WC,LZhu20JSAC_UAVRelay,Sharma20TWC_UAVRelay,Hosseinalipour20TWC_UAVRelay}, multiuser communications \cite{JLyuWCL16_WC,QWuJSAC18_WC,QWuTWC18_multiUAV,LLiuTCom18_MultiUAV}, data collection \cite{Li20IoT_data,Zhan20TCOM_data,You19TWC_data,Wang20IoT_Data,feng2020uavenabled}, secrecy communications\cite{GZhang2019,JYao2020,CZhong2019}, device-to-device (D2D) communications \cite{Wang18JSAC_UAVD2D,Mozaffari16TWC_UAVD2D,Asheralieva19TVT_UAVD2D}, and so on.
Motivated by the advancements of UAV-enabled wireless communications, UAV-enabled WPT has also emerged as a viable solution to resolve the aforementioned technical issues in WPT from a fundamentally new perspective. Different from conventional ETs at fixed locations, low-altitude UAVs can serve as a new type of aerial ETs that can fly flexibly to charge nearby low-power devices efficiently. Specifically,  UAVs can not only enjoy the favorable line-of-sight (LoS) channels with ground devices (GDs) \cite{Hourani18WCL,DMatolak15VTM} but also exploit their fully-controllable mobility to adaptively adjust flight trajectories over time to reduce the transmission distances to GDs based on their real-time locations, thus enhancing the energy transfer efficiency to all GDs significantly and also ensuring their performance fairness effectively. As such, UAV-enabled WPT is particularly appealing for large-scale wireless networks with massive widely distributed GDs, for which the conventional approach of densely deploying fixed-location ETs is very costly if not infeasible. Some promising application scenarios of UAV-enabled WPT such as smart city, maritime communications, wireless sensor networks, and smart factory, are shown in Fig. \ref{fig1}. Furthermore, UAV-enabled WPT can be extended to UAV-enabled WPCN \cite{najmeddint2019energy,wang2018energy,ye2020optimization} and UAV-enabled wireless powered MEC\cite{du2019TVT,Wang2020TNSE,liu2019IoT,zhou2018JSAC,Hu2020TWC}, in which UAVs play multifarious roles such as aerial APs and aerial MEC servers (in addition to ETs) to provide sustainable data access and remote edge computing for GDs, respectively.


To fully reap the benefits of UAV-enabled WPT, how to properly design the UAV trajectories to maximize the energy transfer performance is a new and challenging problem to tackle innovatively. In the most basic case when there is one single UAV wirelessly charging one single GD, it is straightforward that the UAV should fly close to the GD as much as possible to minimize their transmission distance for maximizing the energy transfer efficiency. However, if there are multiple GDs distributed at different locations that are {\it a-prior} known, how to design the UAV's trajectory to balance the transferred energy amounts to different GDs given a finite charging time becomes non-trivial. This is due to the fact that when the UAV flies close to one GD for charging more efficiently, it may have to leave far apart from some other GDs and charge less energy to them, thus leading to a fundamental tradeoff in balancing the transferred energy amounts among different GDs. Moreover, if there are more than one UAV cooperatively charging many GDs in a large-scale network, the joint multi-UAV trajectory design becomes even more challenging.

To tackle the above challenges, there have been a handful of prior works in the literature that investigated the trajectory design for enhancing the energy transfer performance for UAV-enabled WPT when there is only one single UAV \cite{JXuTWC18_WPT,YHu19TCOM_WPT,SKu19_WPT,XMo20WCL_WPT,GeneticUAV,UAVnonlinear}. More specifically, the authors in \cite{JXuTWC18_WPT} first considered the single-UAV-enabled WPT by assuming that the UAV flies at a fixed altitude, in which efficient two-dimensional (2D) UAV trajectory designs are proposed to maximize the sum energy harvested by all GDs or the minimum energy harvested among GDs, subject to the UAV's flight time and speed constraints. Building upon the trajectory design framework in \cite{JXuTWC18_WPT}, the authors in \cite{YHu19TCOM_WPT} proposed the globally optimal one-dimensional (1D) UAV trajectory to maximize the minimum energy harvested among GDs, when all GDs are located on a line, while other works \cite{SKu19_WPT,XMo20WCL_WPT} extended the UAV trajectory design to different setups \cite{GeneticUAV,UAVnonlinear}.
These trajectory designs have been further extended to the UAV-enabled  WPCN in \cite{xie2020common,xie2018throughput,hadzi2019wireless,ye2020optimization} and wireless powered MEC in \cite{du2019TVT,Wang2020TNSE,liu2019IoT,zhou2018JSAC,Hu2020TWC}, where the joint design of UAV trajectory and communication/computation  resource allocations was investigated.


In view of the above existing works, this paper aims to provide a comprehensive and up-to-date tutorial overview on UAV-enabled WPT, with the particular emphasis on how to exploit the UAV trajectory design for optimizing the system performance. This paper is organized as follows.

\begin{itemize}
\item First, Section II considers the single-UAV-enabled WPT scenario with one single UAV wirelessly charging multiple GDs. In this case, we first use a toy example with one single GD to show the benefit of trajectory design, and then present a generic utility maximization problem to maximize the energy amounts transferred to multiple GDs in a fair manner subject to practical UAV flight constraints. To solve the formulated energy maximization problem in this case, we present a general trajectory design framework consisting of three innovative approaches to optimize the trajectory, which are multi-location hovering, successive-hover-and-fly, and time-quantization-based optimization, respectively.
\item Next, Section III considers the multi-UAV-enabled WPT scenario with multiple UAVs cooperatively charging many GDs in a large area, for which a generic utility maximization problem to jointly optimize multiple UAVs' trajectories and their energy transmissions is presented for the first time. While the optimal solution to this problem is still open, we propose two heuristic but efficient schemes based on the principles of UAV swarming and GD clustering, respectively, and design their corresponding multi-UAV trajectories by extending the single-UAV trajectory design solutions.
\item Moreover, Sections IV and V consider two emerging applications of UAV-enabled WPT, namely UAV-enabled WPCN and UAV-enabled wireless powered MEC, respectively. Under properly designed operation protocols for these two applications, we formulate their utility maximization problems to jointly optimize the UAV trajectory and the communication/computation resource allocations, subject to UAV's flight constraints and GD's energy harvesting constraints. By extending the UAV trajectory design framework for WPT, efficient joint UAV trajectory and resource allocation designs are developed.
\item Finally, Section VI provides discussions on the challenging open problems and important future research directions in UAV-enabled WPT, including the issues of  non-linear energy harvesting models, channel state information (CSI) availability, over-the-air computation (AirComp), online trajectory design, and ground vehicles for WPT.
\end{itemize}

It is worth noting that there have been various overview papers on UAV-enabled wireless communications and the corresponding trajectory design/optimization approaches \cite{YZengComM16_Survey,YZongProce19_Survey,MMozaffari19ComS_Survey,Wu2020aa_JSAC20}. However, the trajectory design for UAV-enabled WPT in this paper differs significantly from that for UAV-enabled wireless communications due to the following reasons. First, in UAV-enabled wireless communications, different GDs generally need to communicate with the UAV over orthogonal time-frequency blocks, and accordingly, the UAV should decide its trajectory over time based on the multiple-access scheme and the communicating GDs at each time instant. By contrast, in UAV-enabled WPT, different GDs can simultaneously harvest energy from the same RF signals sent by the UAV. Second, the utility/objective functions for communication versus WPT performance optimization are also different in general. To our best knowledge, a comprehensive tutorial overview on the UAV trajectory design for UAV-enabled WPT is still lacking in the current literature.

{\it Notation:} Scalars are denoted by lower-case letters, vectors by bold-face lower-case letters, and matrices by bold-face upper-case letters. For a vector $\mv{m}$, $\mv{m}^T$ denotes its transpose, and $\| \mv{m} \|$ denotes its Euclidean norm. For a matrix $\mv{M}$, $\mv{M}^{H}$ denotes its hermitian. $\mathbb{E}[\cdot]$ denotes the statistical expectation. $\mathrm{Tr}(\cdot)$ denotes the trace of a square matrix. $\mathrm{diag}(\cdot)$ denotes a diagonal matrix with the argument denoting its main diagonal.

\section{Single-UAV-Enabled WPT}\label{sec:WPT}

This section considers the single-UAV-enabled WPT system, in which one single UAV is dispatched as an aerial ET to wirelessly charge multiple GDs at known locations. In the following, we first provide a toy example with one single GD to show the benefit of UAV trajectory design in enhancing the energy transfer efficiency, then formulate the utility maximization problem to fairly maximize the harvested energy amounts at different GDs, and next present a general trajectory optimization framework, followed by numerical results.

\subsection{Toy Example with One GD}\label{toyexample}
\begin{figure}
\centering
\subfigure[Conventional WPT]
{
    \begin{minipage}{7.5cm}\label{fig2a}
    \centering
    \includegraphics[width=7cm]{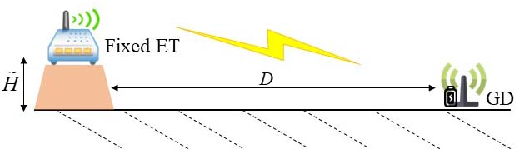}
    \end{minipage}
 }
 \subfigure[UAV-enabled WPT]
{
    \begin{minipage}{7.5cm}\label{fig2b}
    \centering
    \includegraphics[width=6.5cm]{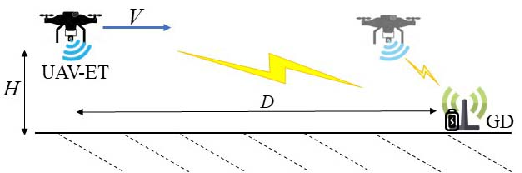}
    \end{minipage}
 }

\caption{A toy example with one single GD.} \label{fig2}
\end{figure}

This subsection considers a toy example with one single GD, as shown in Fig. \ref{fig2}, to show the benefit of UAV-enabled WPT over the conventional WPT with a fixed ET. First, as the benchmark for comparison, we  consider the conventional WPT in the 2D Cartesian coordinate system as shown in Fig. \ref{fig2a}, where the ET and GD are deployed at fixed locations $(0,\tilde H)$ and $(D,0)$, respectively, with distance $\tilde{d}=\sqrt{D^2+\tilde{H}^2}$. In this case, the channel power gain from the ET to the GD is expressed as $\tilde{h}=\tilde{\beta_0}/\tilde{d}^{\tilde{\alpha}}$, where $\tilde{\beta_0}$ denotes the channel power gain at reference distance $d_0=1$~meter (m) and $\tilde{\alpha}$ denotes the path loss exponent with $\tilde{\alpha}\geq2$ in general.
Considering constant transmit power $P$ at the ET and the linear energy harvesting model at the GD for converting the received RF signal into usable energy, the harvested power at the GD is
\begin{align}
	E_{\text{fix}}&=\eta P \tilde{\beta}_0/\tilde{d}^{\tilde{\alpha}} = \frac{\eta P \tilde{\beta_0}}{\left({D^2+\tilde{H}^2}\right)^{\tilde{\alpha}/2}},\label{Efix}
\end{align}
where $0<\eta\le1$ denotes the linear RF-to-direct current (DC) energy conversion efficiency at the GD. It is observed from (\ref{Efix}) that due to the signal path loss, the harvested power $E_{\text{fix}}$ sharply decreases as the distance $D$ becomes large, especially when the path loss exponent $\tilde \alpha$ is high (e.g., when the signal propagation is blocked by obstacles).


Next, we consider the UAV-enabled WPT in a 2D Cartesian coordinate system as shown in Fig. \ref{fig2b}. As the UAV is mobile, we focus on a finite charging period $\mathcal{T}\triangleq(0,T]$ with duration $T$, and denote the UAV's time-varying location as $(x(t),H)$ at time $t\in\mathcal{T}$. Here, $H$ denotes the fixed altitude of the UAV and $H \ge \tilde H$ generally holds due to the UAV's relatively higher altitude than the conventional ET on the ground. Accordingly, the line-of-sight (LoS) link normally exists between the UAV and the GD, and thus we consider the free-space path loss model with path loss exponent $\alpha=2$, as commonly adopted in the UAV literature\cite{wu2018uav,ZengTCom16_Relay_WC}. In order for fair comparison with the conventional WPT in Fig. \ref{fig2a}, we assume that the UAV starts and finishes its charging mission at $x(0) = x(T) = 0$.
%
%
%
%
%
In this case, the channel power gain between the UAV and the GD at any time instant $t\in\mathcal T$ is
\begin{align}
h(x(t))=\beta_0/(d(x(t)))^{2}=\frac{\beta_0}{(x(t)-D)^2+H^2},\label{h}
\end{align}
where $d(x(t))=\sqrt{(x(t)-D)^2+H^2}$ denotes their distance. Accordingly, the harvested power by the GD at time instant $t\in\mathcal T$ is \cite{JXuTWC18_WPT}
\begin{align}
	\bar{E}(x(t))=\frac{\eta  P\beta_0}{(x(t)-D)^2+H^2}.\label{e1}
\end{align}
The total harvested energy by the GD over the entire period $\mathcal T$ is
\begin{align}
	E(\{x(t)\})=\int_0^T\frac{\eta P\beta_0}{(x(t)-D)^2+H^2}\text{d}t.\label{Euav}
\end{align}

By comparing $E(\{x(t)\})/T$ in (\ref{Euav}) and $E_{\text{fix}}$ in (\ref{Efix}), it is observed that as compared to the conventional WPT, the UAV-enabled WPT can enhance the average harvested power at the GD due to the following two main reasons. First, with the relatively higher altitude of the UAV, UAV-enabled WPT enjoys the LoS energy transmission link with lower path loss exponent. Second, the UAV can exploit its mobility via optimizing the trajectory $\{x(t)\}$ to shorten the distance $d(x(t))$ with the GD (e.g., when the UAV hovers exactly above the GD with $x(t)=D$, the distance between the UAV and the GD is minimized as $H$).

To fully exploit such benefits, how to optimize the UAV trajectory $\{x(t)\}$ is crucial, which needs to take into account the UAV's flight constraints in practice. Intuitively, to maximize the harvested energy $E(\{x(t)\})$ at the GD in (\ref{Euav}), the UAV should try its best to fly as close to the GD as possible. In particular, suppose that the UAV is constrained by the maximum flight speed $V_{\max}$. Then the UAV's optimal trajectory design can be obtained as follows by considering two cases.
\begin{itemize}
	\item  When $TV_{\max}\ge D$, the UAV should first fly straightly towards the GD at the maximum flight speed $V_{\max}$ with duration $D/V_{\max}$, then hover above the GD with duration $T-2D/V_{\max}$, and finally fly straightly back to the initial location at speed $V_{\max}$ with duration $D/V_{\max}$.
	\item When $TV_{\max}<D$, the time duration is not sufficient for the UAV to hover above the GD. In this case, the UAV should first fly towards the GD at speed $V_{\max}$ in the first half period, and then fly back in the second half.

\end{itemize}

\begin{figure}
	\centering
	\includegraphics[width=7cm]{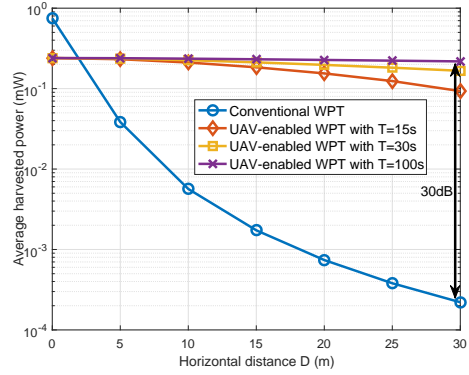}\\
	\caption{The average harvested power versus the GD's horizontal location $D$ under different time duration $T$.}\label{fig3}
\end{figure}

Fig. \ref{fig3} shows the average harvested power by the GD versus its horizontal location $D$, in which $\tilde{\alpha}=3$, $\eta=60\%$, $\tilde{H}=2$~m, $H=5$~m, $P=40$~dBm, and $\beta_0=-30$~dB. It is observed that as the (initial) horizontal distance $D$ between the ET/UAV and the GD increases, for conventional WPT the average harvested power degrades severely, while for UAV-enabled WPT the average harvested power almost remains unchanged (especially when $T$ becomes large). More specifically, when $T=100$~in second (s), UAV-enabled WPT is observed to achieve 30 dB harvested power gain over the conventional WPT. This shows the enormous benefit of exploiting the UAV mobility to combat against the severe signal path loss for enhancing the energy transfer efficiency.

\subsection{Utility Maximization with Multiple GDs}

\begin{figure}
	\centering
	\includegraphics[width=7cm]{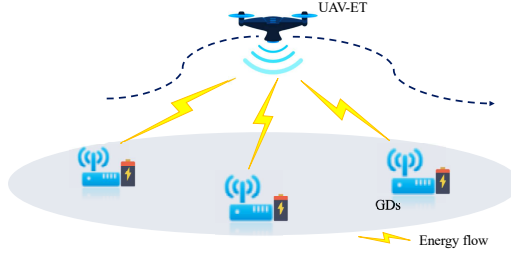}\\
	\caption{Illustration of the UAV-enabled WPT system in the case with multiple GDs.}\label{modelWPT}
\end{figure}

Building upon the insights gained from the simplified case with one single GD in Section \ref{toyexample}, this subsection considers the general single-UAV-enabled WPT in the three-dimensional (3D) Cartesian coordinate system with $K>1$ GDs,
as shown in Fig.~\ref{modelWPT}, in which the UAV is dispatched to serve $K$ GDs over a finite charging period $\mathcal T\triangleq(0,T]$. Let $(x_k,y_k,0)$ denote the location of each GD $k\in\mathcal{K}\triangleq\{1,...,K\}$, where $\boldsymbol{c}_k=(x_k,y_k)$ denotes its horizontal coordinate. Let $(x(t),y(t),H)$ denote the UAV's location at time instant $t\in\mathcal T$, in which $\boldsymbol{u}(t)=(x(t),y(t))$ denotes the time-varying horizontal location to be optimized. Accordingly, the distance between the UAV and GD $k\in\mathcal K$ is $d_k(\mv{u}(t))=\sqrt{\|\mv{u}(t)-\boldsymbol{c}_k\|^2+H^2}$ at time instant $t\in\mathcal T$. Hence, the channel power gain between the UAV and GD $k\in\mathcal{K}$ at time $t\in\mathcal{T}$ is
\begin{align}
	h_k(\boldsymbol{u}(t))=\frac{\beta_0}{\|\mv{u}(t)-\boldsymbol{c}_k\|^2+H^2}.
\end{align}
Similarly as $\bar{E}(x(t))$ in (\ref{e1}), the harvested power by GD $k\in\mathcal{K}$ at time $t\in\mathcal{T}$ is given by
\begin{align}
\hat{E}_k(\mv{u}(t))=\frac{\eta P\beta_0}{\|\mv{u}(t)-\boldsymbol{c}_k\|^2+H^2}.\label{EkWPT1}
\end{align}
As a result, the total harvested energy by GD $k$ over the entire period $\mathcal{T}$ is given by
\begin{align}
E_k(\{\mv{u}(t)\})=\int_0^T\frac{\eta P\beta_0}{\|\mv{u}(t)-\boldsymbol{c}_k\|^2+H^2}\text{d}t.\label{EkWPT}
\end{align}

How to design the UAV trajectory $\{\boldsymbol{u}(t)\}$ to maximize the harvested energy amounts at multiple GDs is a non-trivial problem, as there generally exists a tradeoff in balancing the harvested energy amounts $\{E_k(\{\mv{u}(t)\})\}$ at these distributed GDs. For instance, setting $\boldsymbol{u}(t)=\boldsymbol{c}_k$ can maximize the harvested energy at GD $k$, but may decrease the harvested energy at other GDs that are far apart. To deal with this issue, we introduce a utility function $U(\{\mv{u}(t)\})$ to maximize, which fairly balances the harvested energy amounts at all these GDs, defined as the minimum weighted harvested energy among the $K$ GDs, i.e.,
\begin{align}
U(\{\mv{u}(t)\})=\min\Bigg\{\frac{E_1(\{\mv{u}(t)\})}{a_1},\frac{E_2(\{\mv{u}(t)\})}{a_2},...,\frac{E_K(\{\mv{u}(t)\})}{a_K}\Bigg\}.\label{UtilitySUAV}
\end{align}
Here, $a_k>0$ denotes the constant energy weight of GD $k\in\mathcal{K}$, a larger value of which means that GD $k$ has a higher priority in maximizing its harvested energy.

Based on the utility function $U(\{\mv{u}(t)\})$, we formulate a generic trajectory optimization problem for the single-UAV-enabled WPT system as follows by taking into account practical UAV flight constraints.
\begin{align}
\text{(P1):}&\max_{\{\mv{u}(t)\}}~U(\{\mv{u}(t)\})\nonumber\\
\mathtt{s.t.}&~f_i(\{\mv{u}(t)\})\ge0,\forall i\in\{1,...,I\},\label{P1con}
\end{align}
where $f_i(\cdot)$'s represent constraint functions on the UAV's trajectory and $I$ denotes the number of such constraints. Some widely adopted UAV trajectory constraints are introduced as follows.
\begin{itemize}
\item Flight speed constraints: Suppose that the flight speed of UAV is constrained by a maximum value $V_{\max}$, and thus we have
\begin{align}
	\|\dot{\boldsymbol{u}}(t)\|\le V_{\max},\forall t\in\mathcal T,\label{speedcons}
\end{align}
where $\dot{\boldsymbol{u}}(t)$ denotes the first-order derivation of $\boldsymbol{u}(t)$ with respect to $t$.\footnote{Fixed-wing UAVs are also subject to minimum flight speed constraints in order to keep aloft.}
\item Acceleration constraints: The acceleration of UAVs is also subject to a maximum value $V_{\max}^{AC}$ in practice. This corresponds to second-order constraints on $\boldsymbol{u}(t)$, given by
\begin{align}
	\|\ddot{\boldsymbol{u}}(t)\|\le V^{AC}_{\max},\forall t\in\mathcal T,
\end{align}
where $\ddot{\boldsymbol{u}}(t)$ denotes the second-order derivation of $\boldsymbol{u}(t)$ with respect to $t$.
\item Initial/final location constraints: The UAV normally needs to start/finish its flight mission at certain locations (e.g., at UAV stations for charging). Suppose that the initial and final locations of UAV are $\boldsymbol{u}_I$ and $\boldsymbol{u}_F$, respectively. We thus have
\begin{align}
	\boldsymbol{u}(0) = \boldsymbol{u}_I, \boldsymbol{u}(T) = \boldsymbol{u}_F.
\end{align}

\item Obstacle avoidance constraints: During flight, the UAV should stay away from obstacles in the space. Let $\boldsymbol{o}$ and $d_{\min}$ denote the location of a certain obstacle and the minimum distance requirement from the obstacle, respectively. We thus have
\begin{align}
	\|\boldsymbol{u}(t)-\bold{o}\|\ge d_{\min},\forall t\in\mathcal T.
\end{align}

\end{itemize}
Notice that problem (P1) involves an infinite number of optimization variables over continuous time and the objective function is non-concave with respect to $\{\boldsymbol{u}(t)\}$ in general. Therefore, problem (P1) is a challenging problem to be optimally solved. To tackle this problem, prior works have proposed various efficient algorithms to solve (P1) under different setups and flight constraints (see, e.g., \cite{YHu19TCOM_WPT,SKu19_WPT,XMo20WCL_WPT}). In particular, the authors in \cite{JXuTWC18_WPT} present a framework to optimize the trajectory by considering the flight speed constraints in (\ref{speedcons}), which will be detailed next.

\begin{remark}
	It is worth noticing that the trajectory optimization problem (P1) for the single-UAV-enabled WPT is different from that for the UAV-enabled multiuser wireless communications (see, e.g., \cite{ZengTWC17_WC}). For the UAV-enabled multiuser communications, different GDs will suffer from the inter-user interference in general, and thus the UAV should design its trajectory to not only maximize the received signal power at desired GDs but also minimize the interference power towards non-desired GDs, under properly designed multiple access schemes. By contrast, for the single-UAV-enabled WPT of our interest, different GDs can harvest the wireless energy transferred from the UAV at the same time.  Therefore, the conventional trajectory designs for UAV-enabled multiuser communications (see, e.g., \cite{QWuJSAC18_WC}) are not applicable for the UAV-enabled WPT of our interest.
\end{remark}

\begin{remark}\label{WPT_linear}
	It is also worth noting that, the weighted sum harvested energy can also be adopted as a utility function to balance the harvested energy tradeoff among multiple GDs. However, it has been shown in \cite{JXuTWC18_WPT} that due to the linear relationship between the harvested energy and the channel power gain, the UAV only needs to hover at one single fixed location during the whole charging period to maximize the weighted sum harvested energy of all GDs with their given weights. This solution may cause a severe fairness issue if the weights for GDs are not properly set, as the nearby GDs with the UAV can harvest significantly more energy than far-apart GDs. This issue resembles that with the weighted sum rate maximization adopted in UAV-enabled multiuser communications \cite{UAVsumrate}.
\end{remark}


\subsection{Trajectory Design Framework}\label{UAVsolve}

This subsection introduces the trajectory design framework in \cite{JXuTWC18_WPT} to solve problem (P1). For exploitation, we only consider the speed constraints in (\ref{speedcons}) in problem (P1), and the presented trajectory design framework is extendable to the case with other flight constraints. By introducing an auxiliary variable $E$, problem (P1) with the speed constraints in (\ref{speedcons}) is re-formulated as
\begin{align}
\text{(P1.1):}&\max_{\{\boldsymbol{u}(t)\},E}~E\nonumber\\
\mathtt{s.t.}&~E_k(\{\mv{u}(t)\})/a_k\ge E,\forall k\in\mathcal K\label{P1.1con}\\
&~\|\dot{\boldsymbol{u}}(t)\|\le V_{\max},\forall t\in\mathcal T.\label{P1.1speed}
\end{align}
In the trajectory design framework\cite{JXuTWC18_WPT}, problem (P1.1) is first relaxed by ignoring the UAV flight speed constraints in (\ref{P1.1speed}), for which the optimal multi-location hovering solution is obtained via the Lagrange duality method \cite{boyd2004convex}. Next, building upon the multi-location hovering solution, a heuristic {\it successive hover-and-fly (SHF) trajectory} is presented by leveraging the traveling salesman problem (TSP), such that the UAV can sequentially visit these hovering locations with a shortest path. Finally, the time quantization is implemented to transform the continuous-time trajectory design problem (P1.1) into an equivalent discrete-time trajectory design problem that is solvable via successive convex approximation (SCA)\cite{SCA}, in which the SHF trajectory is adopted as the initial point for iteration. The detailed procedure for solving (P1.1) is given as follows.

\subsubsection{Multi-Location-Hovering Design} First, we consider the relaxed problem by ignoring the UAV flight speed constraints in (\ref{P1.1speed}), which may correspond to the practical case when the time duration $T$ is sufficiently long. The relaxed problem is expressed as
\begin{align}
	\text{(P1.2):}&\max_{\{\boldsymbol{u}(t)\},E}~E\nonumber\\
\mathtt{s.t.}&~\text{(\ref{P1.1con}).}\nonumber
\end{align}
Although problem (P1.2) is still non-convex, it can be shown that it satisfies the so-called time-sharing condition in \cite{1658226}. Consequently, the strong duality holds between problem (P1.2) and its Lagrange dual problem. Therefore, the optimal solution to (P1.2) can be obtained by using the Lagrange duality method. Let $\lambda_k\ge 0,k\in\mathcal{K},$ denote the dual variable associated with the $k$-th constraint in (\ref{P1.1con}). The Lagrangian of (P1.2) is
\begin{align}
	\mathcal L_1(\{\boldsymbol{u}(t)\},\{\lambda_k\},E)=(1-\sum_{k\in\mathcal K}\lambda_k)E+\sum_{k\in\mathcal K}\lambda_kE_k(\{\mv{u}(t)\})/a_k.
\end{align}
The dual function becomes
\begin{align}
	g_1(\{\lambda_k\})=\max_{\{\boldsymbol{u}(t)\},E}~\mathcal L_1(\{\boldsymbol{u}(t)\},\{\lambda_k\},E).\label{dualfunction}
\end{align}
As a result, the dual problem of (P1.2) is
\begin{align}
\text{(D1.2):}&\min_{\{\lambda_k\ge0\}}~g_1(\{\lambda_k\})\nonumber\\
\mathtt{s.t.}&~\sum_{k\in\mathcal{K}}\lambda_k=1,\label{lambda}
\end{align}
where the equality in (\ref{lambda}) must hold in order to ensure $g_1(\{\lambda_k\})$ to be bounded\cite{JXuTWC18_WPT}. Due to the strong duality between (P1.2) and (D1.2), the optimal solution to problem (P1.2) can be obtained by equivalently solving the dual problem (D1.2) as follows.
\begin{itemize}
	\item First, for any given feasible dual variables $\{\lambda_k\}$, solving problem (\ref{dualfunction}) to obtain the dual function $g_1(\{\lambda_k\})$ is equivalent to solving the following problem.
	\begin{align}
		\max_{\{\boldsymbol{u}(t)\}}~\int_0^T\sum_{k\in\mathcal K}\frac{\lambda_k}{a_k}\frac{\eta P\beta_0}{\|\mv{u}(t)-\boldsymbol{c}_k\|^2+H^2}\text{d}t.\label{dualproblem}
	\end{align}
    Notice that in problem (\ref{dualproblem}), the optimization of UAV trajectory $\{\boldsymbol{u}(t)\}$ is independent over time $t$. Therefore, solving problem (\ref{dualproblem}) is equivalent to finding the (hovering) location that maximizes the weighted sum harvested energy of all GDs:
    \begin{align}
    	\max_{\boldsymbol{u}}~\sum_{k\in\mathcal K}\frac{\lambda_k}{a_k}\frac{\eta P\beta_0}{\|\mv{u}-\boldsymbol{c}_k\|^2+H^2}.\label{dualproblem2}
    \end{align}
    The optimal solution to problem (\ref{dualproblem2}) can be obtained by using a 2D exhaustive search over a region $[\underline{x},\underline{y}]\times[\bar{x},\bar{y}]$, where $\underline{x}\triangleq\min\{x_k\}$, $\underline{y}\triangleq\min\{y_k\}$, $\bar{x}\triangleq\max\{x_k\}$, and $\bar{y}\triangleq\max\{y_k\}$. Let $\mv{u}^{\{\lambda_k\}}$ denote the optimal solution to problem (\ref{dualproblem2}), which may not be unique in general. As a result, the dual function $g_1(\{\lambda_k\})$ is obtained.
	\item Next, we find the optimal dual variables to solve the dual problem (D1.2). In general, the dual function $g_1(\{\lambda_k\})$ is always convex but generally non-differentiable. With $g_1(\{\lambda_k\})$ obtained, we solve the dual problem by subgradient-based methods such as the ellipsoid method \cite{boyd2004convex}. Let $\{\lambda_k^{\text{opt}}\}$ denote the optimal dual solution, and $\mv{u}^{\{\lambda^{\text{opt}}_k\}}_\omega,\omega\in\{1,...,\Omega\}$, denote the optimal hovering location solution to problem (\ref{dualproblem2}) under $\{\lambda_k^{\text{opt}}\}$, where $\Omega\ge1$ is the number of the (non-unique) hovering location solutions.
	\item Finally, we construct the optimal primal solution to problem (P1.2) based on the obtained $\mv{u}^{\{\lambda^{\text{opt}}_k\}}_\omega$ from the optimal dual solution. As the solution $\mv{u}^{\{\lambda^{\text{opt}}_k\}}_\omega$ is generally non-unique (i.e., $\Omega>1$ generally holds), proper time sharing among these hovering locations is essential, such that the UAV should hover above the $\Omega$ locations over time to maximize the minimum weighted harvested energy. Let $\tau_\omega\ge 0$ denote the time-sharing factor or equivalently the hovering duration at the $\omega$-th location, $\omega\in\{1,...,\Omega\}$, which can be obtained via solving the following linear programming (LP).
	\begin{align}
		\max_{\{\tau_\omega\ge 0\}}~&E\nonumber\\
		\mathtt{s.t.}~&\sum_{\omega=1}^{\Omega}\tau_\omega E_k(\{\mv{u}^{\{\lambda^{\text{opt}}_k\}}_\omega\})/a_k\ge E,\forall k\in\mathcal K\nonumber\\
		&\sum_{\omega=1}^\Omega \tau_{\omega}=T.\label{LP}
	\end{align}
    Suppose that the optimal solution to the LP is $\{\tau_\omega^*\}$. Then the optimal solution to problem (P1.2) is obtained, which corresponds to that the UAV successively hovers among the $\Omega$ locations $\{\mv{u}^{\{\lambda^{\text{opt}}_k\}}_\omega\}$, each with duration $\tau_\omega^*$. This is thus called the multi-location-hovering solution. Note that the performance achieved by this solution is the upper bound of the optimal value of problem (P1.1) with the speed constraints in (\ref{P1.1speed}).
\end{itemize}

\subsubsection{SHF Trajectory Design} Next, building upon the multi-location-hovering solution to the relaxed problem (P1.2), a heuristic SHF trajectory design is presented by taking into account the UAV's flight speed constraints in (\ref{P1.1speed}). With the SHF trajectory, the UAV sequentially visits and hovers above each of the obtained hovering locations $\mv{u}^{\{\lambda^{\text{opt}}_k\}}_\omega$ and then flies among them straightly at the maximum flight speed $V_{\max}$. How to properly order the visited hovering locations and optimize the hovering duration at each location is crucial.

Towards this end, we first determine the UAV's traveling path to visit all the optimal hovering locations with minimum flying distance. It is shown in \cite{JXuTWC18_WPT} that the traveling path minimization problem can be transformed into a modified TSP without initial and final locations,\footnote{In standard TSP, the salesman needs to find the optimal traveling path with minimum traveling distance to visit a number of cities and return to the origin city\cite{TSP}. } and thus can be solved efficiently by using some well-established methods such as integer programming\cite{TSPinteger}.
Next, the hovering duration optimization can be formulated as an LP similar to problem (\ref{LP}) by further taking into account the harvested energy during the straight flight (see \cite[(25)]{JXuTWC18_WPT}). By combining the flight path and the hovering durations, the SHF trajectory design for solving (P1) is completed.


\subsubsection{Time Quantization Based Optimization}
Besides the SHF trajectory, time quantization is another widely adopted approach to obtain the trajectory based on convex optimization techniques. In particular, the whole mission duration is discretized into a finite number of $N$ time slots, each with equal duration $\delta=T/N$, where the duration $\delta$ is sufficiently small such that the UAV location is assumed to be approximately unchanged during each slot, denoted as $ {\boldsymbol{u}}[n]=\boldsymbol{u}(n\delta)$. Accordingly, the minimum harvested energy maximization problem in (P1.1) with continuous-time trajectory can be reformulated as the following problem (P1.3) with discrete-time waypoints.
\begin{align}
	\text{(P1.3):}&\max_{\{{\boldsymbol{u}}[n]\},E}~E\nonumber\\
\mathtt{s.t.}&~\frac{1}{Ta_k}\sum_{n=1}^N\frac{\eta \delta P\beta_0}{\|{\mv{u}}[n]-\boldsymbol{c}_k\|^2+H^2}\ge E,\forall k\in\mathcal K\label{P1.3con}\\
&~\|{\boldsymbol{u}}[n]-{\boldsymbol{u}}[n+1]\|\le \delta V_{\max},\forall n\in\{1,...,N-1\},\label{P1.3speed}
\end{align}
where (\ref{P1.3speed}) is the time-quantized UAV flight speed constraints.
Although problem (P1.3) is still non-convex, we can obtain a high-quality solution by utilizing the SCA technique, which approximates the non-convex problem into a sequence of convex problems (see \cite[(35)]{JXuTWC18_WPT}) that can be efficiently solved by convex optimization techniques such as CVX \cite{CVX}. It is shown that by properly choosing the approximate functions, the convergence of the iterative SCA algorithm can be ensured \cite{JXuTWC18_WPT}.
Note that the performance of SCA-based algorithm critically depends on the initial point of iteration, and the proposed SHF trajectory design can serve as a high-quality initial point.

\subsection{Numerical Results}\label{WPTsimulation}
\begin{figure}
  \centering
  \includegraphics[width=7cm]{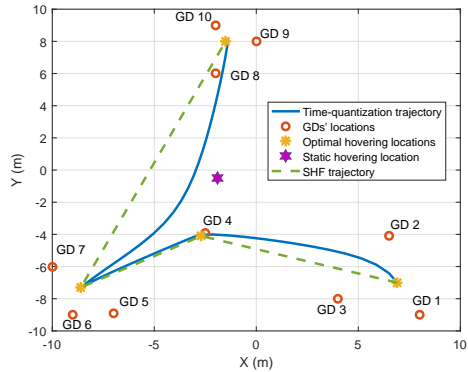}\\
  \caption{Simulation system setup and trajectory design for single-UAV-enabled WPT with $K=10$ GDs.}\label{fig4}
\end{figure}

\begin{figure}
	\centering
	\includegraphics[width=7cm]{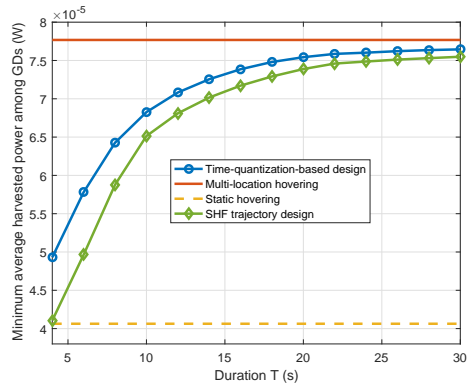}\\
	\caption{The minimum average harvested power of GDs versus the UAV mission duration $T$.}\label{fig:WPTperformance}
\end{figure}

This subsection presents numerical results to evaluate the performance of the above three trajectory design approaches, as compared to the benchmarking static hovering scheme without exploiting the UAV's mobility over time. In the static hovering scheme, the UAV hovers at one single location over the whole mission period to maximize the minimum weighted harvested energy, i.e., $\max_{\boldsymbol{u}}\min_{k\in\mathcal{K}}\{\hat E_k(\{\mv{u}\})/a_k\}$, for which the optimal static hovering solution can be obtained via a 2D exhaustive search.

In this simulation, we consider a single-UAV-enabled WPT system with $K=10$ GDs that are randomly distributed within a 2D area of $20\times20$~m$^2$, as shown in Fig. \ref{fig4}, where the parameters are set same as those for Fig. \ref{fig3}, and the maximum speed of the UAV is $V_{\max}=5$~m/s. In order to balance the harvested energy among all the GDs, we consider that $a_k=1,\forall k\in\mathcal K$.

Fig. \ref{fig4} shows the obtained trajectories based on the proposed approaches and the static hovering, where $T=50$~s. It is observed that there are 4 optimal hovering locations in the multi-location-hovering design, which are close to GDs 8-10, GDs 5-7, GD 4, and GDs 1-3, respectively, in order to charge them efficiently. It is also observed that the time-quantization-based trajectory deviates from the straight line of the SHF trajectory to maximize the energy transfer efficiency during the flight.

Fig. \ref{fig:WPTperformance} shows the minimum average harvested power  among all the GDs versus the UAV mission duration $T$. It is observed that the SHF and time-quantization-based trajectory designs significantly outperform the static hovering scheme and the performance gain becomes more significant when $T$ increases. The time-quantization-based trajectory is observed to outperform the SHF trajectory. Furthermore, when $T$ is sufficiently large, the SHF and time-quantization-based trajectory designs are observed to perform close to the performance upper bound achieved by the multi-location-hovering solution with the UAV's flight speed constraints ignored.




\section{Multi-UAV-Enabled WPT}\label{SEC_MultiUAV}
\begin{figure}
  \centering
  \includegraphics[width=7cm]{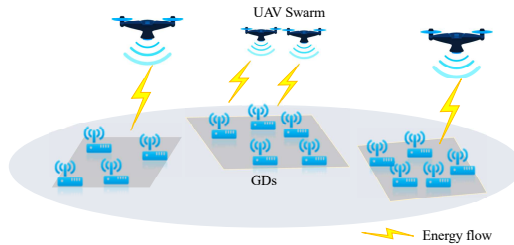}\\
  \caption{Illustration of the multi-UAV-enabled WPT.}\label{modelMUAV}
\end{figure}
The preceding section studied the trajectory design for single-UAV-enabled WPT, which, however, may not work well when there are many GDs in a large area. Therefore, this section considers the general multi-UAV-enabled WPT system as shown in Fig. \ref{modelMUAV}, in which multiple UAVs are dispatched to cooperatively charge the GDs. Different from the trajectory design for single-UAV-enabled WPT, multiple UAVs need to cooperatively design their trajectories jointly with their energy transmission to maximize the energy transfer performance.

Suppose that there are a set, $\mathcal M=\{1,...,M\}$, of single-antenna UAVs cooperatively charging a set, $\mathcal K=\{1,...,K\}$, of GDs over the charging period $\mathcal T=(0,T]$. At time instant $t\in\mathcal{T}$, let $\mv{u}_m(t), m\in\mathcal{M}$, denote the horizontal location of UAV $m$. Supposing that all UAVs stay at the same altitude $H$, the distance between UAV $m$ and GD $k$ is $d_{k,m}(\mv{u}_m(t))=\sqrt{\|\boldsymbol{u}_m(t)-\boldsymbol{c}_k\|^2+H^2}$. Under the LoS channel assumption, we express the channel coefficient between UAV $m$ and GD $k$ as follows, by considering the free-space path loss together with a random phase.
\begin{align}
\bar{h}_{k,m}(\mv{u}_m(t))) = \sqrt{\frac{\beta_0}{\|\mv{u}_m(t)-\boldsymbol{c}_k\|^2+H^2}} e^{j \theta_{k,m}(t)},\label{MUAVchannel}
\end{align}
where $j=\sqrt{-1}$, and $\theta_{k,m}(t)$ denotes the random phase that is assumed to be a uniformly distributed random variable in $[0,2\pi)$. Notice that similar channel models in (\ref{MUAVchannel}) have been widely adopted in the UAV-enabled wireless communication literature (see, e.g.,\cite{LLiuTCom18_MultiUAV}), in order to capture the fact that the channel phase changes much more rapidly in practice than the free-space path loss.\footnote{Note that in the single-UAV-enabled WPT case, the fluctuation of channel phases does not affect the harvested energy amount at each GD, which is thus omitted in Section \ref{sec:WPT}.} By collecting the wireless channels from different UAVs, we denote $\bar{\mv{h}}_k(t)=[\bar{h}_{k,1}(t),\bar{h}_{k,2}(t)...,\bar{h}_{k,M}(t)]^T$ as the channel vector from the $M$ UAVs to each GD $k\in\mathcal{K}$.

Next, we consider the cooperative energy transmission at the $M$ UAVs. Let ${s}_m(t)$ denote the transmit energy signal at each UAV $m\in\mathcal M$, and ${\bm s}(t)=[{s}_1(t),\cdots,{s}_m(t)]^T$ denote the collective transmit signals by the $M$ UAVs. Accordingly, we denote $ {\bm S}(t)=\mathbb{E}[{\bm s}(t){\bm s}^H(t)]$ as the transmit energy covariance matrix. Notice that if ${\bm S}(t)$ is a diagonal matrix, then the $M$ UAVs design their respective transmit energy signals independently, while if ${\bm S}(t)$ is of rank-one, then the $M$ UAVs use cooperative beamforming to send one energy beam with common energy signals. In this case, the total harvested energy of GD $k\in\mathcal{K}$ over the entire period $\mathcal T$ is
\begin{align}
\tilde{E}_k(\{\mv{u}_m(t),\mv{S}(t)\})=\int_0^T \mathbb{E}\left[\eta\bar{\mv{h}}_k^H(t)\mv{S}(t)\bar{\mv{h}}_k(t)\right]\text{d}t,\label{energyMUAV}
\end{align}
where the expectation is taken with respect to the randomness in the channel phases.

It is observed in (\ref{energyMUAV}) that the harvested energy amounts at the $K$ GDs are related to both the $M$ UAVs' trajectories $\{\boldsymbol{u}_m(t)\}$ and their transmit covariance matrices $\{{\bm S}(t)\}$. Therefore, in order to maximize the harvested energy at the $K$ GDs, we should jointly optimize both $\{\boldsymbol{u}_m(t)\}$ and $\{{\bm S}(t)\}$. Similarly as in (\ref{UtilitySUAV}) for single-UAV-enabled WPT, we define the utility function for multi-UAV-enabled WPT as
\begin{align}
	U(\{\mv{u}_m(t),\mv{S}(t)\})=\min\Bigg\{\frac{E_1(\{\mv{u}_m(t),\mv{S}(t)\})}{a_1},\frac{E_2(\{\mv{u}_m(t),\mv{S}(t)\})}{a_2},...,\frac{E_K(\{\mv{u}_m(t),\mv{S}(t)\})}{a_K}\Bigg\}.
\end{align}
As a result, we have the generic utility maximization problem for multi-UAV-enabled WPT as
\begin{align}
\text{(P2):}\max_{\{\boldsymbol{u}_m(t),\boldsymbol{S}(t)\}}~&U(\{\mv{u}_m(t),\mv{S}(t)\})\nonumber\\
\mathtt{s.t.}&~f_{i,m}(\{\boldsymbol{u}_m(t)\})\ge0,\forall i\in\{1,...,I\},m\in\mathcal{M}\\
&\bar{f}_{l}(\{\boldsymbol{u}_m(t)\})\ge0,\forall l\in\{1,...,L\}\\
&~p_j(\boldsymbol{S}(t))\ge0,\forall j\in\{1,...,J\},\label{perUAVpowercon}
\end{align}
where $f_{i,m}(\cdot)$'s denote the flight constraints for each individual UAV $m$ similarly as in (\ref{P1con}), and $\bar{f}_{l}(\cdot)$'s denote the joint flight constraints for multiple UAVs with $L$ representing the number of such constraints. For instance, the collision avoidance constraints to ensure safe flight for multiple UAVs can be given by
\begin{align}
	\|\boldsymbol{u}_i(t)-\boldsymbol{u}_j(t)\|\le d_{\min},\forall i,j\in\mathcal M,i\neq j,t\in\mathcal{T},\label{collisioncon}
\end{align}
where $d_{\min}$ denotes the safety distance between any two different UAVs.
Moreover, $p_j(\cdot)$'s represent the constraints on the transmit signals at UAVs with $J$ representing the number of such constraints. For example, supposing that each UAV $m$ is subject to a maximum transmit power $P_m$, we have the individual power constraints as $\mathrm{Tr}(\boldsymbol{C}_m\boldsymbol{S}(t))\le P_m,\forall t\in\mathcal T,m\in\mathcal M$, where $\boldsymbol{C}_m$ denotes a matrix with only the element in the $m$-th column and $m$-th row being 1 and all the other elements being 0. Assuming that the sum power of all the $M$ UAVs are constrained by a maximum value $P_{\text{sum}}$, we have the sum-power constraints as $\mathrm{Tr}(\boldsymbol{S}(t))\le P_{\text{sum}},\forall t\in\mathcal T$. In the following, we consider the individual power constraints for UAVs with $P_m=P,\forall m\in\mathcal{M}$.

It is worth noting that problem (P2) is much more challenging to be optimally solved than problem (P1) due to the involvement of multiple UAVs' trajectories and their transmit covariance matrices, and such joint optimization may require information sharing among different UAVs.
While finding the optimal solution to problem (P2)
is still open and has not been investigated in the literature yet, in the next two subsections, we propose two heuristic designs based on the principles of UAV swarming and GD clustering, respectively, by extending the trajectory design framework for single-UAV-enabled WPT in Section \ref{sec:WPT}.

\subsection{UAV Swarming with TDMA based Beamforming}\label{swarm}
First, we consider the UAV swarming based design, in which all UAVs are formed into a swarm to fly following the same speed and orientation. Suppose that UAV 1 acts as the swarm head with the time-varying location being $\boldsymbol{u}_1(t)=(x_1(t),y_1(t))$ at time $t\in\mathcal{T}$. Furthermore, assuming that the UAVs' formation is fixed over time, the location of UAV $i\in\{2,...,M\}$ is given as $\boldsymbol{u}_i(t)=\boldsymbol{u}_1(t)+\boldsymbol{b}_i$, where $\boldsymbol{b}_i$ denotes its relative coordinate with respect to $\boldsymbol{u}_1$. For example, supposing that there are four UAVs that are formed into a square with edge length $d_{\min}$, then the coordinates of UAVs 2, 3, and 4 at time $t$ can be expressed as $\boldsymbol{u}_2(t)=(x_1(t),y_1(t)-d_{\min})$, $\boldsymbol{u}_3(t)=(x_1(t)-d_{\min},y_1(t)-d_{\min})$, and $\boldsymbol{u}_4(t)=(x_1(t)-d_{\min},y_1(t))$, respectively.
In this case, the design of multiple UAVs' trajectories in problem (P2) is reduced into the design of  trajectory $\{\boldsymbol{u}_1(t)\}$ for the swarm head UAV 1 only.\footnote{Notice that the UAV swarming with such a fixed formulation can also be viewed as a single UAV with multiple antennas. Therefore, the problem formulation and developed designs are also applicable for the single-UAV-enabled WPT design with multiple antennas.}



Next, we consider the design of transmit covariance matrices $\{{\bm S}(t)\}$. While finding the optimal $\{{\bm S}(t)\}$ is non-trivial, we consider the TDMA-based beamforming design, in which all UAVs cooperatively send one energy beam towards one GD at each time. At time instant $t\in\mathcal T$, let $\tau_{E,k}(t)\in\{0,1\}$ denote an energy beamforming indicator, where $\tau_{E,k}(t)=1$ denotes that the $M$ UAVs design the energy beam towards GD $k$ at time $t$, and $\tau_{E,k}(t)=0$ otherwise. Therefore, we have $\sum_{k\in\mathcal{K}}\tau_{E,k}(t)=1,\forall t\in\mathcal{T}$. Accordingly, under $\tau_{E,k}(t)=1$, each UAV $m\in\mathcal M$ sets the transmitted signal as $\sqrt{P}e^{j\phi_{m,k}(t)}s^E$, where $s^E$ denotes the common energy signal with unit power, and $\phi_{m,k}(t)=-\theta_{k,m}(t)$, such that the energy signals will be coherently combined at GD $k$.\footnote{Notice that the cooperative beamforming design requires the UAVs to have symbol-level time synchronization, such that the received signals can be coherently combined at the desired GD.} In this case, the harvested power by GD $k$ is given by
\begin{align}
\tilde{E}_k(\{&\boldsymbol{u}_m(t)\})=\eta \mathbb{E}\left[\left|\sum_{m=1}^M \sqrt{P}|\bar{h}_{k,m}(t)| s^E\right|^2\right]= \eta P\left(\sum_{m=1}^M\sqrt{ \frac{\beta_0}{\lVert{\boldsymbol{u}}_m(t)-\boldsymbol{c}_k\rVert^2+H^2}}\right)^2,
\end{align}
and that by any other GD $\bar{k}\in\mathcal K\setminus\{k\}$ with $\tau_{E,\bar{k}}=0$ (without coherent combining) is
\begin{align}
{E}'_{\bar k}(\{\boldsymbol{u}_m(t)\})&=\eta \mathbb{E}\left[\left|\left(\sum_{m=1}^M \sqrt{P}\bar{h}_{\bar k,m}(t) e^{j (\theta_{k,m}(t) - \theta_{\bar k,m}(t))}\right) s^E\right|^2\right] =\eta P\sum_{m=1}^M \frac{\beta_0}{\|{\boldsymbol{u}}_m(t)-\boldsymbol{c}_{\bar k}\|^2+H^2}.
\end{align}
Therefore, the harvested power by GD $k$ at time $t\in\mathcal{T}$ is given by
\begin{align}
	\tilde{E}_k^{\text{har}}(\{\boldsymbol{u}_m(t)\},\tau_{E,k}(t))&=\left(\tau_{E,k}(t)\tilde{E}_k(\{\boldsymbol{u}_m(t)\})+(1-\tau_{E, k}(t)){E}'_{k}(\{\boldsymbol{u}_m(t)\})\right).
\end{align}
The total harvested energy by GD $k$ over the entire period $\mathcal{T}$ is given by
\begin{align}
\tilde{E}_k^{\text{tot}}(\{\boldsymbol{u}_m(t),\tau_{E,k}(t)\})&=\ \int_0^T\tilde{E}_k^{\text{har}}(\boldsymbol{u}_m(t),\tau_{E,k}(t))\text{d}t.
\end{align}

Under the UAV swarming and TDMA-based beamforming designs and by considering the UAV flight speed constraints only (similarly as in Section \ref{sec:WPT}), the utility maximization problem (P2) is simplified as
\begin{align}
\text{(P2.1):}\max \limits_{ \{\boldsymbol{u}_1(t),\tau_{E,k}(t)\}} \min\limits_{k\in\mathcal K } ~& \tilde{E}_k^{\text{tot}}(\{\boldsymbol{u}_m(t),\tau_{E,k}(t)\})/a_k\notag\\
\mathtt{s.t.} ~~~~~~~~~~&\|\dot{\boldsymbol{u}}_1(t)\|\le V_{\max},\forall t\in\mathcal T\label{Swarm_Pro1}\\
~~~~~~~~~~&\tau_{E,k}(t)\in\{0,1\},\forall k\in\mathcal{K},t\in\mathcal{T}\label{tau1}\\
~~~~~~~~~~&\sum_{k\in\mathcal{K}}\tau_{E,k}(t)=1,\forall t\in\mathcal{T},\label{tau2}
\end{align}
which can be equivalently re-written as follows by introducing an auxiliary variable $\tilde E$.
\begin{align}
\text{(P2.2):}\max \limits_{ \{\boldsymbol{u}_1(t),\tau_{E,k}(t)\},\tilde E} & \tilde E \notag\\
\mathtt{s.t.} ~~~~& \tilde{E}_k^{\text{tot}}(\{\boldsymbol{u}_m(t),\tau_{E,k}(t)\})/a_k\ge \tilde E,\forall k\in\mathcal{K}\label{Swarm_Pro2_energy}\\
&(\ref{Swarm_Pro1}),~(\ref{tau1}),~\text{and}~(\ref{tau2}),\notag
\end{align}
in which only the trajectory design of the swarm head (UAV 1) needs to be optimized, jointly with the beamforming indicators $\{\tau_{E,k}(t)\}$.
It is observed that problem (P2.1)/(P2.2) has similar structures as (P1.1)/(P1.2), except the newly involved beamforming indicators $\{\tau_{E,k}(t)\}$. Motivated by this, in the following we extend the trajectory design framework in Section \ref{UAVsolve} to solve problem (P2.1)/(P2.2).

\subsubsection{Multi-Location-Hovering Design}
Firstly, we consider the following relaxed problem by ignoring the UAV flight speed constraints in (\ref{Swarm_Pro1}).
\begin{align}
	\text{(P2.3):}\max \limits_{ \{\boldsymbol{u}_1(t),\tau_{E,k}(t)\},\tilde E} ~ &\tilde E \nonumber\\
	\mathtt{s.t.} ~~~~&(\ref{tau1}),~(\ref{tau2}),~\text{and}~(\ref{Swarm_Pro2_energy}).\notag
\end{align}
Similarly as for problem (P1.2), although problem (P2.3) is still non-convex, the strong duality holds between (P2.3) and its dual problem. Therefore, the optimal solution to (P2.3) can be obtained by using the Lagrange duality method. Let $\lambda_k\ge0,k\in\mathcal{K}$ denote the dual variable associated with the $k$-th constraints in (\ref{Swarm_Pro2_energy}). The partial Lagrangian of (P2.3) is
\begin{align}
	\mathcal{L}_2(\{\boldsymbol{u}_m(t),\tau_{E,k}(t),\lambda_k\},\tilde E)=(1-\sum_{k\in\mathcal{K}}\lambda_k)\tilde E+\ \int_0^T\sum_{k\in\mathcal{K}}\lambda_k\tilde{E}_k^{\text{har}}(\boldsymbol{u}_m(t),\tau_{E,k}(t))\text{d}t.
\end{align}
The dual function becomes
\begin{align}
	g_2(\{\lambda_k\})=\max_{\{\boldsymbol{u}_m(t),\tau_{E,k}(t)\},\tilde E}~&\mathcal{L}_2(\{\boldsymbol{u}_m(t),\tau_{E,k}(t),\lambda_k\},\tilde E)\nonumber\\
	\mathtt{s.t.}&~(\ref{tau1}),~(\ref{tau2}).\label{MUAVdual}
\end{align}
As a result, the dual problem of (P2.3) is
\begin{align}
	\text{(D2.3):}\min_{\{\lambda_k\ge0\}}~&g_2(\{\lambda_k\})\nonumber\\
	\mathtt{s.t.}~&\sum_{k\in\mathcal K}\lambda_k=1.
\end{align}
The optimal solution to (P2.3) can be obtained by equivalently solving (D2.3). First, for any given feasible dual variables, the dual function $g_2(\{\lambda_k\})$ in (\ref{MUAVdual}) can be obtained by solving the following problem.
\begin{align}
	\max_{\{\boldsymbol{u}_1(t),\tau_{E,k}(t)\}}~&\int_0^T\sum_{k\in\mathcal{K}}\lambda_k\left(\tau_{E,k}(t)\tilde{E}_k(\{\boldsymbol{u}_m(t)\})+(1-\tau_{E, k}(t)){E}'_{k}(\{\boldsymbol{u}_m(t)\})\right)\text{d}t\nonumber\\
	\mathtt{s.t.}&~(\ref{tau1}),~(\ref{tau2}).\label{MUAV33}
\end{align}
As problem (\ref{MUAV33}) consists of infinite number of identical sub-problems for different time $t$, solving problem (\ref{MUAV33}) is equivalent to
\begin{align}
	\max_{\boldsymbol{u}_1,\{\tau_{E,k}\}}~&\sum_{k\in\mathcal{K}}\lambda_k\eta P\left(\tau_{E,k}\left(\sum_{m=1}^M\sqrt{ \frac{\beta_0}{\lVert{\boldsymbol{u}}_m-\boldsymbol{c}_k\rVert^2+H^2}}\right)^2+\sum_{m=1}^M \frac{(1-\tau_{E, k})\beta_0}{\|{\boldsymbol{u}}_m(t)-\boldsymbol{c}_{\bar k}\|^2+H^2}\right)\label{MUAVdual42}\\
	\mathtt{s.t.}~&\tau_{E,k}\in\{0,1\},\forall k\in\mathcal{K}\label{MUAVdual42con1}\\
	&\sum_{k\in\mathcal{K}}\tau_{E,k}=1,\label{MUAVdual42con2}
\end{align}
for each time $t\in\mathcal{T}$. Note that for the optimization of the beamforming indicators $\{\tau_{E,k}\}$ in (\ref{MUAVdual42}), there are $K$ feasible choices satisfying constraints (\ref{MUAVdual42con1}) and (\ref{MUAVdual42con2}), each with $\tau_{E,k}=1,k\in\mathcal{K}$, and accordingly $\tau_{E,\bar k}=0, \forall \bar k\in\mathcal K,\bar k\neq k$. Under $\tau_{E,k}=1$, problem (\ref{MUAVdual42}) is equivalent to finding the optimal hovering location via solving the following problem:
\begin{align}
	\max_{\boldsymbol{u}_1}~\lambda_k\eta P\left(\sum_{m=1}^M\sqrt{ \frac{\beta_0}{\lVert{\boldsymbol{u}}_m-\boldsymbol{c}_k\rVert^2+H^2}}\right)^2+\sum_{\bar k\in\mathcal{K}\setminus\{k\}}\lambda_{\bar k}\eta P\sum_{m=1}^M \frac{\beta_0}{\|{\boldsymbol{u}}_m-\boldsymbol{c}_{\bar k}\|^2+H^2},\label{MUAVdual43}
\end{align}
which can be solved by using a 2D exhaustive search. Hence, by comparing the obtained optimal values under the $K$ choices for problem (\ref{MUAVdual43}), the optimal solution to problem (\ref{MUAVdual42}) can be obtained as $\boldsymbol{u}_1^{\{\lambda_k\}}$ and $\tau_{E,k}^{\{\lambda_k\}}$, and accordingly the dual function $g_2(\{\lambda_k\})$ is also obtained. Note that if any two of the $K$ optimal values are equal, then the optimal solution to problem (\ref{MUAVdual42}) can be non-unique, which leads to multiple hovering locations corresponding to energy beamforming design towards different GDs.

Next, we use subgradient-based methods to solve the dual problem (D2.3), where $\{\lambda_k^{\text{opt}}\}$ denote the accordingly obtained optimal dual solution. Under $\{\lambda_k^{\text{opt}}\}$, let $\mathcal{K}^{\text{opt}}$ denote the set of GD $k$'s such that $\tau_{E,k}^{\{\lambda_k^{\text{opt}\}}}=1$ is one (non-unique) solution to (\ref{MUAVdual42}) for any $k\in\mathcal{K}^{\text{opt}}$, and  $\boldsymbol{u}_{1,k,\omega}^{\{\lambda_k^{\text{opt}}\}},\omega\in\{1,...,\Omega_k\},$ denote the optimal hovering location solution to problem (\ref{MUAVdual43}) under any $k\in\mathcal{K}^{\text{opt}}$, with $\Omega_k$ denoting the corresponding number of optimal hovering locations. In this case, there are a total number of $\sum_{k\in\mathcal{K}^{\text{opt}}}\Omega_k$ optimal hovering locations. 

Finally, based on these optimal hovering locations, we need to time share among them via solving an LP similarly to problem (\ref{LP}). By implementing this, the optimal multi-location-hovering solution to primal problem (P2.3) is found, in which the UAVs need to hover among the $\sum_{k\in\mathcal{K}^{\text{opt}}}\Omega_k$ locations over time, over each of which the UAVs need to accordingly design the beamforming towards the corresponding GD.

\subsubsection{SHF Trajectory Design}
Next, by taking into account the UAV flight speed constraints, the SHF trajectory design can be constructed via using the TSP algorithm, in which the UAV swarm visits and hovers above each of the above obtained optimal hovering locations and flies among them straightly at the maximum flight speed $V_{\max}$. Under the flight path obtained by TSP, we still need to optimize the hovering duration at each location and the beamforming indicators over time. However, the optimization of beamforming indicators involves an infinite number of variables. To tackle this issue, we can quantize the entire UAV charging period into a finite number of equal-duration time slots, in each of which the UAV swarm is assumed to be stay at fixed locations. Furthermore, we can divide each time slot into $K$ sub-slots with variable durations, and in each sub-slot $k$, all UAVs design their energy beamforming towards GD $k$. In this case, the optimization of hovering durations and beamforming indicators corresponds to optimizing the durations of sub-slots over time, which can be found via solving an LP.

\subsubsection{Time Quantization Based Optimization}

Finally, we can directly adopt time quantization to transform problem (P2.2) with continuous-time variables into an equivalent optimization problem with discrete-time variables. We discretize the whole charging period $\mathcal T$ into a set $\mathcal{N}\triangleq\{1,...N\}$ of time slots each with equal duration $\delta=T/N$, and further divide each slot to $K$ sub-slots, during each of which the UAVs perform cooperative energy beamforming towards GD $k$. At time slot $n\in\mathcal{N}$, let $\boldsymbol{u}_m[n]=\boldsymbol{u}(n\delta),m\in\mathcal{M}$, denote the UAV $m$'s location and $\tau_{E,k}[n],k\in\mathcal{K}$ denote the duration of sub-slot $k$, where we have $\sum_{k\in\mathcal{K}}\tau_{E,k}[n]=\delta$. Accordingly, by introducing two sets of auxiliary variables of $\{\alpha_{k,m}[n]\}$ and $\{ A_k[n]\}$, problem (P2.2) is reduced to
\begin{align}
\text{(P2.3):}\max \limits_{ \substack{\{\boldsymbol{u}_1[n],\tau_{E,k}[n]\ge0\},\tilde E, \\
\{\alpha_k[n]\ge0\},\{ A [n]\ge 0\}}} & \tilde E \nonumber\\
\mathtt{s.t.} ~~~~& \frac{\eta P}{Ta_k} \sum_{n=1}^N \left( \tau_{E,k}[n]A_k[n]+(\delta-\tau_{E,k}[n]) \sum_{m=1}^M \Phi_{k,m}({\boldsymbol{u}}_m[n])\right) \ge \tilde E,\forall k\in\mathcal{K}\label{Swarm_Pro3_energy}\\
& A_k[n]\le \Psi(\alpha_{k,m}[n]), \forall k\in\mathcal K,n\in\mathcal N\label{Swarm_Pro3_A}\\
&\alpha_{k,m}^2[n]\le \Phi_k({\boldsymbol{u}}_m[n]), \forall k\in\mathcal K,m\in\mathcal{M},n\in\mathcal N\label{Swarm_Pro3_alpha}\\
&\sum_{k\in\mathcal{K}}\tau_{E,k}[n]=\delta,\forall n\in\mathcal{N}\label{Beamindicator47}\\
&(\ref{Swarm_Pro1}),\notag
\end{align}
where $\Phi_{k,m}({\boldsymbol{u}}_m[n])=\frac{\beta_0}{\lVert{\boldsymbol{u}}_m[n]-\boldsymbol{c}_k\rVert^2+H^2}$ and $\Psi_k(\alpha_{k,m}[n])=\left(\sum_{m=1}^M \alpha_{k,m}[n] \right)^2$.
However, the constraints in \eqref{Swarm_Pro3_energy}-\eqref{Swarm_Pro3_alpha} are still non-convex due to the coupling of the time allocation and UAV trajectory. Therefore, we apply an alternating-optimization-based method to solve this problem iteratively, in which we alternately optimize the time allocation and the UAV trajectory by assuming the other to be given.

Under given trajectory $\{\boldsymbol{u}_1[n]\}$ (and equivalently $\{\boldsymbol{u}_m[n]\}$), the optimization of the time allocation $\{\tau_{E,k}[n]\}$ corresponds to an LP, which can be efficiently solved by CVX. Therefore, we only need to focus on optimizing the UAV trajectory.

Under any given time allocation $\{\tau_{E,k}[n]\}$, the UAV trajectory optimization is still non-convex, due to the non-convex constraints \eqref{Swarm_Pro3_energy}-\eqref{Swarm_Pro3_alpha}.
To deal with these non-convex constraints, we update the
UAV swarm head's trajectory $\{\boldsymbol{u}_1[n]\}$ (equivalently $\{\boldsymbol{u}_m[n]\}$) and $\{\alpha_{k,m}[n]\}$ in an iterative manner by applying the SCA method.
Let $\{\boldsymbol{u}^{(i)}_1[n]\}$ and $\{\alpha^{(i)}_{k,m}[n]\}$ denote the local points at the $i$-th iteration. By taking the Taylor expansion at any point, the lower bounds of $ \Psi_k(\alpha_{k,m}[n])$ and $\Phi_{k,m}({\boldsymbol{u}}_m[n])$ are given by
\begin{align}
	\Psi_k(\alpha_{k,m}[n])&\ge \Psi_k(\alpha^{(i)}_{k,m}[n])+\Psi'_k(\alpha^{(i)}_{k,m}[n])(\alpha_{k,m}[n]-\alpha^{(i)}_{k,m}[n]))\nonumber\\
	&\triangleq \Psi^{\text{low}}_k(\alpha_{k,m}[n]),\\
\Phi_{k,m}({\boldsymbol{u}}_m[n])&\ge \Phi_{k,m}({\boldsymbol{u}}^{(i)}_m[n])+\Phi'_{k,m}({\boldsymbol{u}}^{(i)}_m[n])(\boldsymbol{u}_m[n]-\boldsymbol{u}^{(i)}_m[n])\nonumber\\
&\triangleq \Phi^{\text{low}}_k({\boldsymbol{u}}_m[n]).
\end{align}
By replacing $\Psi_k(\alpha_{k,m}[n])$ and $\Phi_{k,m}({\boldsymbol{u}}_m[n])$ as $\Psi^{\text{low}}_k(\alpha_{k,m}[n])$ and $\Phi^{\text{low}}_k({\boldsymbol{u}}_m[n])$, we have
\begin{align}
\max \limits_{ \substack{\{\boldsymbol{u}_1[n]\},\tilde E\ge 0, \notag\\
\{\alpha_{k,m}[n]\ge0\},\{ A_k [n]\ge 0\}}} & \tilde E \notag\\
\mathtt{s.t.} ~~~~& \frac{\eta P}{a_kT} \sum_{n=1}^N \left( \tau_{E,k}[n]A_k[n]+(\delta-\tau_{E,k}[n]) \sum_{m=1}^M \Phi^{\text{low}}_k({\boldsymbol{u}}_m[n])\right)  \ge \tilde E,\forall k\in\mathcal{K}\notag\\
& A_k[n]\le \Psi^{\text{low}}_k(\alpha_{k,m}[n]), \forall k\in\mathcal K,n\in\mathcal N\notag\\
&\alpha_{k,m}^2[n]\le \Phi^{\text{low}}_k({\boldsymbol{u}}_m[n]), \forall m\in\mathcal{M}\notag\\
&(\ref{Swarm_Pro1}),~(\ref{Beamindicator47}).\label{UAVtrapro}
\end{align}
The above problem is convex and thus can be solved by the CVX tool. Accordingly, an optimized solution to problem (P2.3) is obtained.

\subsection{GD clustering}
Next, we consider another intuitive design based on GD clustering, which groups different GDs into a number of clusters, and allows different UAVs to design their trajectories in a distributed manner, each covering one cluster. In particular, we separate $K$ GDs into $M$ clusters each served by one UAV. Suppose that the set of GDs in cluster $m\in\mathcal{M}$ is denoted by $\mathcal{K}_m$ and the number of GDs in each cluster $m$ is denoted by $K_m=|\mathcal{K}_m|$, where $\cup_{m\in\mathcal{M}}\mathcal{K}_m=\mathcal{K}$ and $\sum_{m\in\mathcal M}K_m=K$. Accordingly, the whole area is divided into $M$ non-overlapping sub-areas, and each UAV is dispatched to serve the GDs within one corresponding sub-area. Furthermore, different from the UAV swarming design with cooperative transmit energy beamforming used, we consider that different UAVs send independent signals with the transmit covariance matrices being ${\bm S}(t)=\mathrm{diag}(P,...,P)$. For notational convenience, let $\boldsymbol{c}_{k,m}=(x_{k,m},y_{k,m}), k\in\mathcal{K}_m,m\in\mathcal M$ denote the horizontal location of GD $k$ in cluster $m$, and denote the distance between UAV $m$ and GD $k$ in cluster $m$ as $d_{k,m}(\boldsymbol{u}_m(t))=\sqrt{\|\boldsymbol{u}_m(t)-\boldsymbol{c}_{k,m}\|^2+H^2}$. In this case, the total harvested energy by GD $k\in\mathcal{K}_m$ in cluster $m\in\mathcal M$ is given by
\begin{align}
E_{k,m}^{\text{clu}}(\{\boldsymbol{u}_m(t)\})&=\int_0^T\sum_{i\in\mathcal M}\frac{\eta P \beta_0}{\|\boldsymbol{u}_i(t)-\boldsymbol{c}_{k,m}\|^2+H^2}\text{d}t.\label{clusterenergy}
\end{align}
Hence, the utility maximization problem (P2) is reduced as
\begin{align}
	\text{(P2.4):}\max_{\{ \boldsymbol{u}_m(t) \}}~&\min_{k\in\mathcal{K}_m,m\in\mathcal{M}}\{E_{k,m}^{\text{clu}}(\{\boldsymbol{u}_m(t)\})\}\nonumber\\
	\mathtt{s.t.}~&\|\dot{\boldsymbol{u}}_m(t)\|\le V_{\max},\forall m\in\mathcal M,t\in\mathcal T\nonumber\\
	&(\ref{collisioncon}).
\end{align}
Notice that as each UAV covers one sub-area, the collision avoidance constraints between different UAVs can be automatically satisfied. Furthermore, we suppose that the distance between UAV $m$ and any non-associated GD $k,k\notin\mathcal{K}_m$, is generally long, and as a result, we can safely omit the harvested energy from nearby UAVs in (\ref{clusterenergy}) and approximate $E_{k,m}^{\text{clu}}(\{\boldsymbol{u}_m(t)\})$ as
\begin{align}
	E_{k,m}^{\text{clu}}(\{\boldsymbol{u}_m(t)\})\approx\int_0^T\frac{\eta P \beta_0}{\|\boldsymbol{u}_m(t)-\boldsymbol{c}_{k,m}\|^2+H^2}\text{d}t.
\end{align}
As a result, the multi-UAV trajectory design problem (P2.4) can be decomposed into $M$ single-UAV trajectory design problems as follows, which can thus be solved by the trajectory design framework in Section \ref{UAVsolve}.
\begin{align}
	\max_{\{ \boldsymbol{u}_m(t) \}}~&\min_{k\in\mathcal{K}_m}\int_0^T\frac{\eta P \beta_0}{\|\boldsymbol{u}_m(t)-\boldsymbol{c}_{k,m}\|^2+H^2}\text{d}t\nonumber\\
	\mathtt{s.t.}~&\|\dot{\boldsymbol{u}}_m(t)\|\le V_{\max},\forall t\in\mathcal T,
\end{align}
$m\in\mathcal{M}.$ It is worth noting that how to optimally group these GDs is still an open problem that is difficult to solve. Intuitively, we can assign nearby GDs into one cluster via clustering methods such as K-Means clustering. In this paper, we simply divide the whole area are into $M$ sub-areas in a uniform manner, as will be illustrated next.

\subsection{Numerical Results}

\begin{figure}
	\centering
	\includegraphics[width=7cm]{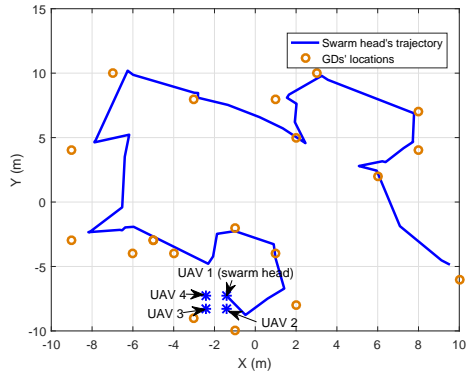}\\
	\caption{Optimized UAV trajectories of multi-UAV-enabled WPT under the UAV swarming design.}\label{swamtra}
\end{figure}

\begin{figure}
	\centering
	\includegraphics[width=7cm]{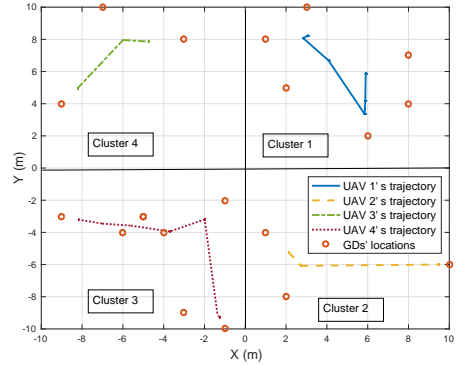}\\
	\caption{Optimized UAV trajectories of multi-UAV-enabled WPT under the GD clustering design.}\label{grouptra}
\end{figure}

In this subsection, we present numerical results to evaluate the performance of the proposed UAV-swarming and GD-clustering schemes as compared to the single-UAV-enabled WPT system with single antenna. In the simulation, we consider that $M=4$ UAVs are dispatched to serve $K=20$ GDs that are randomly distributed within a 2D area of $20\times20$~m$^2$ as shown in Figs. \ref{swamtra} and \ref{grouptra}. The system parameters are same as those for Fig. \ref{fig3}. In particular, for GD clustering, we simply separate the 2D area into four equal-size square sub-areas as shown in Fig. \ref{grouptra}. 


Figs. \ref{swamtra} and \ref{grouptra} show the optimized UAVs' trajectories via time quantization under UAV swarming and GD-clustering designs, respectively. For UAV swarming, it is observed in Fig. \ref{swamtra} that the UAV swarm flies close to each GD sequentially to maximize the energy beamforming gain for that GD (under TDMA beamforming). For GD clustering, it is observed in Fig. \ref{grouptra} that the UAV in each sub-area flies following a similar trajectory as that in Fig. \ref{fig4}.

\begin{figure}
	\centering
	\includegraphics[width=7cm]{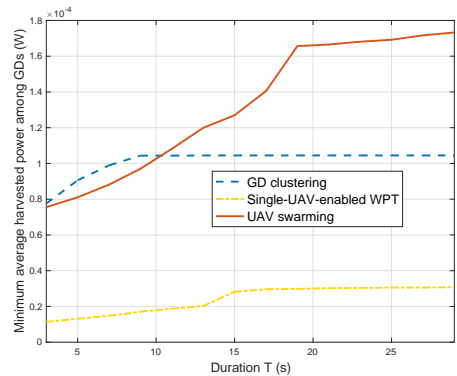}\\
	\caption{The minimum average harvested power among all GDs versus mission duration $T$ under different UAV cooperation schemes.}\label{MUAVDper}
\end{figure}

Fig. \ref{MUAVDper} shows the minimum average harvested power among all GDs versus the UAV charging duration $T$. It is observed that the multi-UAV-enabled WPT (via both UAV swarming and GD clustering) considerably outperforms the single-UAV counterpart. Specifically, when $T$ is small (e.g., $t<10$~s), the GD clustering design outperforms the UAV swarming, as the UAV swarm design generally needs longer durations to properly visit all GDs to charge them efficiently. By contrast, when $T$ becomes large (e.g., $T>11$~s), the UAV swarming surpasses the GD clustering design, due to the exploitation of cooperative transmit energy beamforming.



\subsection{Extensions}
As an initial attempt, this section presented two efficient trajectory designs for multi-UAV-enabled WPT based on UAV swarming and GD clustering, respectively. There are still various potential extensions and open problems that are worth pursuing in future work, which are briefly discussed in the following.
\begin{itemize}
	\item For UAV swarming, we considered simplified TDMA beamforming design and assumed fixed UAV formations over time. Generally speaking, using more advanced cooperative transmit energy beamforming and enabling more adaptive UAV formation are expected to achieve better energy transfer performance. For instance, when the GDs are randomly distributed over space, sending multiple energy beams at each time instant may further enhance the energy transmission efficiency.
	\item For GD clustering, we optimized the trajectory of each UAV independently by omitting the transferred energy from nearby UAVs. In practice, we can jointly optimize the multiple UAVs' trajectories by directly solving problem (P2.4). Besides, how to properly cluster these GDs will be crucial for performance optimization, which is still an open problem.
	\item Furthermore, motivated by the performance comparison in Fig. \ref{MUAVDper}, the ideas of UAV swarming and GD clustering can be combined to further enhance the efficiency of multi-UAV-enabled WPT. For instance, we can group GDs into several clusters each served by a sub-group of UAVs, where for each cluster  UAV-swarming-based cooperative energy beamforming design can be adopted. In this case, we need to decide the number of clusters and the number of UAVs in each cluster, based on which we also need to cluster GDs and jointly design the UAVs' trajectories and energy transmission strategies. These problems are all challenging to solve, for which the performance-complexity tradeoff should be considered properly.
\end{itemize}

\section{UAV-Enabled WPCN}\label{SEC_WPCN}
\begin{figure}
  \centering
  \includegraphics[width=7cm]{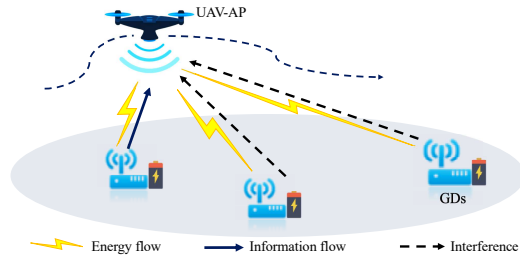}\\
  \caption{Illustration of the UAV-enabled WPCN.}\label{modelWPCN}
\end{figure}

Building upon the UAV-enabled WPT in the preceding two sections, this section considers the UAV-enabled WPCN, in which
UAVs are dispatched as arial hybrid APs to not only wirelessly charge GDs but also collect information from them. By exploiting the UAVs' mobility via trajectory design, the UAV-enabled WPCN is able to resolve the doubly near-far issue faced in the conventional WPCN with APs fixed on the ground. Nevertheless, due to the involvement of both downlink WPT and uplink wireless information transmission (WIT) as well as the newly introduced energy harvesting constraints at GDs, it is a critical task in UAV-enabled WPCN to optimize the UAV trajectory design jointly with the communication resource allocations. To gain essential insights on such joint optimization, in the following, we focus on the single-UAV-enabled WPCN system over a finite mission period $\mathcal{T}=(0,T]$, in which a single UAV flies in the sky to wirelessly charge a set of distributed GDs, and each GD uses the harvested energy to send information back to the UAV.

\subsection{Transmission Protocol}
For the UAV-enabled WPCN, we need to properly design the transmission protocol to share the limited time-frequency resources for both multiuser WIT in the uplink and WPT in the downlink.
In such transmission protocol, frequency division duplex (FDD) and time division duplex (TDD) are two duplexing schemes that are widely adopted, in which the downlink WPT and uplink WIT are implemented over orthogonal frequency and time resources, respectively.\footnote{In-band full duplex (see, e.g., \cite{ye2020optimization,fullduplex}) can also be implemented, in which the UAV's uplink information reception may suffer from strong self-interference from its downlink energy transmission. In this case, effective self-interference cancellation becomes essential.} Furthermore, to support multiuser WIT in the uplink, TDMA, FDMA, OFDMA, and even NOMA\cite{NOMADingZhiguo} can be implemented as the uplink multiple access schemes. Generally speaking, under different duplexing and uplink multiple access schemes, the UAV trajectory design and resource allocations can be different.

In particular, this section considers the transmission protocol with TDD and TDMA as in \cite{xie2018throughput}, in which the downlink WPT and uplink multiuser WIT are implemented over different time instants but in the same frequency bands. Notice that the consideration with TDD and TDMA is particularly appealing for UAV-enabled WPCN, as the UAV can properly adjust its locations over time for efficient WIT and WPT, respectively. Also notice that the presented designs in this section are extendible to other cases when different multiple access and/or duplexing schemes are employed.

Under TDD and TDMA, we define
\begin{align}
	\tau_E(t)\in\{0,1\},\tau_k(t)\in\{0,1\},\forall k\in\mathcal K,\label{taucon1}
\end{align}
as indicators to denote the transmission mode of the UAV-enabled WPCN at time instant $t\in\mathcal{T}$. Here, $\tau_E(t)=1$ means the downlink WPT mode, in which the UAV broadcasts wireless energy to the GDs at time $t$. Similarly, $\tau_k(t)=1$ corresponds to the uplink WIT mode with GD $k$, in which GD $k$ sends information to the UAV at time $t$. As the downlink WPT and uplink WIT with different GDs cannot be implemented at the same time, we have
\begin{align}
	\tau_E(t)+\sum_{k\in\mathcal K}\tau_k(t)\le 1,\forall t\in\mathcal{T}.\label{taucon2}
\end{align}

First, we consider the downlink WPT mode with $\tau_E(t)=1$ and $\tau_k(t)=0,\forall k\in\mathcal{K}$. The UAV adopts constant transmit power $P$. Similarly as in (\ref{EkWPT1}) for single-UAV-enabled WPT, the harvested power by each GD $k\in\mathcal{K}$ at time $t\in\mathcal{T}$ is
\begin{align}
	\bar{E}_k(\mv{u}(t),\tau_E(t))=\tau_E(t)\eta  P h_k(\mv{u}(t)).
\end{align}
Accordingly, the total harvested energy by GD $k$ over the entire period $\mathcal{T}$ is given by
\begin{align}
\bar{E}_k^{\text{tot}}(\{\mv{u}(t),\tau_E(t)\})=\int_0^T  \tau_E(t)\eta P h_k(\mv{u}(t))\text{d}t.\label{WPCN_RP}
\end{align}

Next, we consider the WIT mode with GD $k$, where $\tau_k(t)=1$, and $\tau_E(t)=0,\tau_j(t)=0,\forall j\in\mathcal{K},j\neq k$. Supposing that the transmit power at GD $k$ is $Q_k(t)$ that can be adaptively adjusted over time, then the achievable data rate of GD $k\in\mathcal{K}$ at time $t\in\mathcal{T}$ is
\begin{align}
	r_k(\mv{u}(t),Q_k(t),\tau_k(t))= \tau_k(t)\log_2\left(1+\frac{Q_k(t)h_k(\mv{u}(t))}{\sigma^2}\right),
\end{align}
where $\sigma^2$ is the noise power at the information receiver of the UAV. Accordingly, the average data-rate throughput of GD $k$ over the entire period $\mathcal{T}$ is given by
\begin{align}
R_k(\{\mv{u}(t),Q_k(t),\tau_k(t)\})=\frac{1}{T}\int_0^T \tau_k(t)\log_2\left(1+\frac{Q_k(t)h_k(\mv{u}(t))}{\sigma^2}\right)\text{d}t.
\end{align}

\subsection{Generic Rate Maximization Problem Formulation}
In UAV-enabled WPCN, our objective is to maximize the communication performance of uplink WIT while ensuring the sustainable operation of GDs. In this case, new energy harvesting constraints are imposed at GDs. Suppose that $E_k^{\text{initial}}\ge 0$ denotes the initial energy stored at GD $k$. Then we have the following energy harvesting constraints for GDs to avoid the energy outage\cite{energyoutage}:
\begin{align}
\int_0^t\tau_k(\tilde t)Q_k(\tilde t)\text{d}\tilde t&\le\int_0^t\tau_E(\tilde t) \eta  P h_k(\mv{u}(\tilde t))\text{d}\tilde t+E_k^{\text{initial}},\forall k\in\mathcal K, t\in\mathcal T,\label{energycau}
\end{align}
such that the accumulative energy consumed by each GD until any time $t\in\mathcal{T}$ does not exceed the initial energy plus the energy accumulatively harvested from the UAV by that time. Notice that if $E_k^{\text{initial}}$ is sufficiently large (or equivalently, each GD has a sufficiently large energy storage), the energy harvesting constraints in (\ref{energycau}) can also be replaced by
\begin{align}
\int_0^T \tau_k(t)Q_k(t)\text{d}t&\le \int_0^T\tau_E(t) \eta  P h_k(\mv{u}(t))\text{d}t,\forall k\in\mathcal K,
\end{align}
such that we only need to ensure that the total energy consumption over the entire period $\mathcal{T}$ at each GD does not exceed the totally harvested energy at that GD, for its sustainable operation.

In order to balance the communication rates at different GDs in a fair manner, we define the utility function for UAV-enabled WPCN as the minimum weighted uplink date-rate throughput, i.e.,
\begin{align}
	&\bar{U}(\{\mv{u}(t),Q_k(t),\tau_E(t),\tau_k(t)\})\nonumber\\
	&=\min\Bigg\{\frac{R_1(\{\mv{u}(t),Q_k(t),\tau_1(t)\})}{\bar a_1},\frac{R_2(\{\mv{u}(t),Q_k(t),\tau_2(t)\})}{\bar a_2},...,\frac{R_K(\{\mv{u}(t),Q_k(t),\tau_K(t)\})}{\bar a_K}\Bigg\},
\end{align}
where $\bar a_k$ denotes the rate weight for each GD $k$.
Accordingly, we formulate the generic uplink average data-rate throughput maximization problem as problem (P3) in the following, in which both the UAV trajectory $\{\boldsymbol{u}(t)\}$ and the resource allocation $\{Q_k(t),\tau_E(t),\tau_k(t)\}$ need to be jointly optimized.
\begin{align}
\text{(P3):}&\max_{\{\boldsymbol{u}(t),Q_k(t),\tau_E(t),\tau_k(t)\}}~\bar{U}(\{\mv{u}(t),\tau_E(t),\tau_k(t),Q_k(t)\})\nonumber\\
\mathtt{s.t.}&~f_i(\{\mv{u}(t)\})\ge0,\forall i\in\{1,...,I\}\label{con50}\\
&\int_0^T \tau_k(t)Q_k(t)\text{d}t\le \int_0^T \tau_E(t)\eta  P h_k(\mv{u}(t))\text{d}t,\forall k\in\mathcal K\label{P3con}\\
&(\ref{taucon1}),~(\ref{taucon2}).\nonumber
\end{align}

Note that compared with problem (P1) with trajectory design only, problem (P3) involves new variables $\{Q_k(t),\tau_E(t),\tau_k(t)\}$ for resource allocation and new energy harvesting constraints in (\ref{P3con}). Furthermore, the UAV trajectory $\{\boldsymbol{u}(t)\}$ and the resource allocation $\{Q_k(t),\tau_E(t),\tau_k(t)\}$ are coupled at both the rate and energy functions in the objective as well as constraints. Therefore, problem (P3) is more challenging to be optimally solved than problem (P1).



\subsection{Joint UAV Trajectory and Resource Allocation Design Framework}\label{Pro3_solu}

In this subsection, we extend the trajectory design framework in Section \ref{UAVsolve} to obtain efficient solutions to problem (P3) \cite{xie2018throughput}. For illustration, we only consider the UAV's flight speed constraints in (\ref{speedcons}), i.e., we consider (P3) by replacing constraint (\ref{con50}) by (\ref{speedcons})  \cite{xie2018throughput}.

\subsubsection{Multi-Location-Hovering Design}
By considering the case with sufficiently long time duration, we relax problem (P3) as follows by omitting the UAV's flight constraints.
\begin{align}
	\text{(P3.1):}&\max_{\{\boldsymbol{u}(t),Q_k(t),\tau_E(t),\tau_k(t)\},R}~R\nonumber\\
\mathtt{s.t.}&~R_k(\{\mv{u}(t),Q_k(t),\tau_k(t)\})/\bar{a}_k\ge R,\forall k\in\mathcal K\label{P3.1con}\\
&(\ref{taucon1}),~(\ref{taucon2}),~(\ref{P3con}),\nonumber
\end{align}
where $R$ is an auxiliary variable. Similarly as for problem (P1.2), strong duality holds between the non-convex problem (P3.1) and its dual problem. Therefore, (P3.1) can be optimally solved by using the Lagrange duality method. Let $\lambda_k\ge0$ and $\mu_k\ge0,k\in\mathcal{K}$, denote the dual variables associated to the $k$-th constraints in (\ref{P3.1con}) and (\ref{P3con}), respectively. The partial Lagrangian is given as
\begin{align}
	\mathcal L_3(\{\boldsymbol{u}(t),Q_k(t),\tau_E(t),\tau_k(t),\lambda_k,\mu_k\},R)	=(1-\sum_{k\in\mathcal K}\lambda_k)R+\sum_{k\in\mathcal K}\lambda_kR_k(\{\mv{u}(t),Q_k(t),\tau_k(t)\})/\bar{a}_k\nonumber\\
+\sum_{k\in\mathcal K}\mu_k\left(\bar{E}_k^{\text{tot}}(\{\boldsymbol{u}(t),\tau_E(t)\})-\int_0^T\tau_k(t)Q_k(t)\text{d}t\right).
\end{align}
The dual function is
\begin{align}
g_3(\{\lambda_k,\mu_k\})=\max_{\{\boldsymbol{u}(t),Q_k(t),\tau_E(t),\tau_k(t)\},R}~&\mathcal L_3(\{\boldsymbol{u}(t),Q_k(t),\tau_E(t),\tau_k(t),\lambda_k,\mu_k\},R)\nonumber\\
\mathtt{s.t.}~&(\ref{taucon1}),~(\ref{taucon2}).\label{pro57}
\end{align}
Accordingly, the optimal solution to (P3.1) can be obtained by solving the following dual problem.
\begin{align}
	\text{(D3.1):}\min_{\{\lambda_k\ge0\},\{\mu_k\ge0\}}~&g_3(\{\lambda_k,\mu_k\})\nonumber\\
	\mathtt{s.t.}~&\sum_{k\in\mathcal K}\lambda_k=1.
\end{align}
For given feasible $\{\lambda_k,\mu_k\}$, the dual function $g_3(\{\lambda_k,\mu_k\})$ can be obtained by solving problem (\ref{pro57}).
It can be shown that problem (\ref{pro57}) can be decomposed into infinite number of sub-problems for different time $t$. By dropping the time index $t$, solving problem (\ref{pro57}) is equivalent to solving the following sub-problems for any time $t$.
\begin{align}
	\max_{\{Q_k,\tau_k\},\boldsymbol{u},\tau_E}~&\sum_{k\in\mathcal K}\left(\tau_k\left(\frac{\lambda_k}{\bar a_k}\log_2\left(1+\frac{Q_k\beta_0/\sigma^2}{\|\boldsymbol{u}-\boldsymbol{c}_k\|^2+H^2}\right)-\sum_{k\in\mathcal K}\mu_kQ_k\right)+\frac{\tau_E\mu_k\eta P\beta_0/\sigma^2}{\|\boldsymbol{u}-\boldsymbol{c}_k\|^2+H^2}\right)\nonumber\\
	\mathtt{s.t.}~&\tau_E\in\{0,1\},\tau_k\in\{0,1\},\forall k\in\mathcal{K}\nonumber\\
	&\sum_{k\in\mathcal{K}}\tau_k+\tau_E=1.\label{subpro1}
\end{align}
There are $K+1$ feasible choices for $\tau_E$ and $\{\tau_k\}$, due to the constraints in (\ref{subpro1}). Thus, we exhaustively search over the $K+1$ choices to find the optimal solution to (\ref{subpro1}). First, when GD $k$ transmits information to the UAV with $\tau_k=1,\tau_E=0,\tau_j=0,\forall j\in\mathcal K, j\neq k$, we have  $\boldsymbol{u}_k^*=\boldsymbol{c}_k$ (i.e., the UAV hovers exactly above GD $k$), $Q_{\bar k}^*=0,\forall \bar k\in\mathcal{K},\bar k\neq k,$ and $Q_k^*$ can be obtained by checking the first-order derivative of the resultant objective function. Second, when the system operates in downlink WPT mode with $\tau_E=1,\tau_k=0,\forall k\in\mathcal{K}$, we have $Q_k^*=0,\forall k\in\mathcal{K}$, and the UAV's optimal hovering locations can be obtained by using a 2D exhaustive search similarly to that for single-UAV-enabled WPT, which is generally non-unique. By comparing these $K+1$ optimal values, problem (\ref{pro57}) is solved.

Next, after optimally solving the dual problem (D3.1) via subgradient-based methods, we can accordingly obtain a set of optimal hovering locations and the corresponding communication resource allocation. By properly time-sharing these hovering locations, we can construct the optimal multi-location-hovering solution to (P3.1). It is shown in \cite{xie2018throughput} that the UAV should hover above two different sets of locations, one set among different GDs for downlink WPT (with $Q_k^*=0,\forall k\in\mathcal{K}$) and the other set exactly above each GD $k$ for uplink WIT (with $Q_k>0$ and $Q_{\bar k}^*=0,\forall \bar k\in\mathcal{K},\bar k\neq k$). Note that the performance achieved by such multi-location-hovering solution serves as a performance upper bound for the optimal value of problem (P3) with the UAV's flight speed constraints in (\ref{speedcons}).

\subsubsection{SHF Trajectory Design}
With UAV's flight constraints considered, we can construct the SHF trajectory based on TSP, in which the UAV sequentially visits the two sets of hovering locations at the maximum flight speed $V_{\max}$ with a shortest traveling path. Under this flight path, we still need to optimize the hovering duration at each location and resource allocations over time. Towards this end, we quantize the charging period into a finite number of equal-duration slots, and further divide each slot into $K+1$ sub-slots with variable durations for each GD's uplink WIT and the downlink WPT, respectively. In this case, the optimization of hovering durations and resource allocation can be formulated as a convex optimization problem, which can be solved via CVX\cite{xie2018throughput}.

\subsubsection{Time Quantization Based Optimization}
Furthermore, problem (P3) with continuous-time variables can be transformed into an equivalent problem with discrete-time variables via time quantization. Specifically, the whole mission period $\mathcal{T}$ is quantized into a set, $\mathcal{N}\triangleq\{1,...,N\}$, of time slots, each with equal duration $\delta=T/N$. Particularly, suppose that the $K+1$ transmission modes time share within each slots, such that we define $\tau_E[n]$ and $\tau_k[n],k\in\mathcal{K}$ as the time-sharing duration of downlink WPT and that of uplink WIT of GD $k$ at time slot $n\in\mathcal{N}$, respectively. We thus have
\begin{align}
	\sum_{k\in\mathcal{K}}\tau_k[n]+\tau_E[n]=\delta,\forall n\in\mathcal{N}.\label{tausumcon}
\end{align}
Let $Q_k[n]$ denote GD $k$'s transmit power and $\boldsymbol{u}[n]$ denote the UAV's location at time slot $n\in\mathcal{N}$. As a result, problem (P3) with continuous-time variables can be reformulated as the following problem.
\begin{align}
	\text{(P3.2):}&\max_{\{\boldsymbol{u}[n],Q[n],\tau_E[n],\tau_k[n]\}}~\min_{k\in\mathcal{K}}\left\{\frac{1}{T\bar a_k}\sum_{n\in\mathcal{N}} \tau_k[n]\log_2\left(1+\frac{Q_k[n]/\sigma^2}{\|\boldsymbol{u}[n]-\boldsymbol{c}_k\|^2+H^2}\right)\right\}\nonumber\\
	\mathtt{s.t.}&~\sum_{n\in\mathcal{N}}\tau_k[n]Q_k[n]\le \sum_{n\in\mathcal{N}} \frac{\tau_E[n]\eta  P}{\|\boldsymbol{u}[n]-\boldsymbol{c}_k\|^2+H^2},\forall k\in\mathcal K\\
	&(\ref{P1.3speed}),~(\ref{tausumcon}).\nonumber
\end{align}
Although (P3.2) is still challenging to be solved, we can optimize the UAV's trajectory and resource allocations in an alternating manner together with SCA techniques towards a locally optimal solution\cite{xie2018throughput}, in which the SHF trajectory design can be utilized as the initial point for iteration.

\subsection{Numerical Results}
\begin{figure}
  \centering
  \includegraphics[width=7cm]{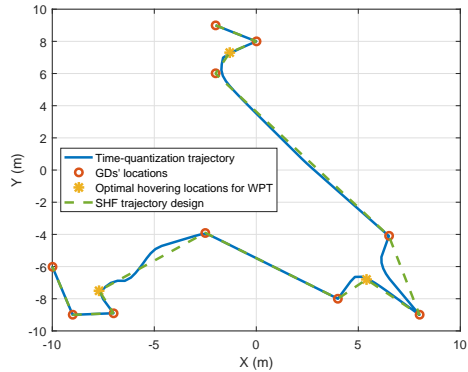}\\
  \caption{Simulation system setup for UAV-enabled WPCN.}\label{fig5}
\end{figure}

\begin{figure}
	\centering
	\includegraphics[width=7cm]{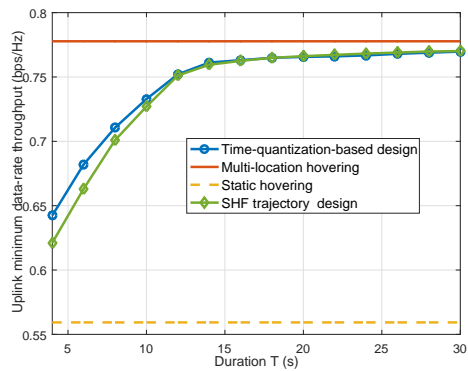}\\
	\caption{The uplink minimum data-rate throughput versus the UAV mission duration $T$.}\label{fig:WPCNperformance}
\end{figure}

In this subsection, we present numerical results to validate the efficiency of the above approaches as compared to the benchmarking static hovering scheme, in which the UAV hovers at one (optimized) location over the whole mission period.


We consider a UAV-enabled WPCN with $K=10$ GDs, in which the simulation parameters are the same as those in UAV-enabled WPT in Section \ref{WPTsimulation}. Fig. \ref{fig5} shows the optimal hovering locations, the trajectory obtained by time-quantization-based approach, and the static hovering location. It is observed that there are 3 optimal hovering locations for WPT, and 10 hovering locations each exactly above one GD for WIT.

Fig. \ref{fig:WPCNperformance} shows the uplink minimum data-rate throughput among all GDs versus the UAV mission duration $T$. It is observed that the time-quantization-based design significantly outperforms the static hovering scheme and the performance gain becomes more evident when $T$ increases. Furthermore, when $T$ is sufficiently large, the time-quantization-based design approaches the performance upper bound achieved by the multi-location hovering without UAV's flight speed constraints.

\subsection{Extensions}

So far, this section presented joint UAV trajectory design and communication resource allocation for the single-UAV-enabled WPCN with one single UAV serving in the dual role of ET and AP. In the literature, there have also been works that extended the design framework to other setups\cite{hu2020aoi,wu2019energy,najmeddint2019energy,wang2018energy,wang2019joint,tang2020minimum,wu2019minimum,ye2020optimization,wang2018resource,hadzi2020uav,li2019throughput,park2019uav}, as briefly discussed in the following.


\subsubsection{Dual-UAV-Enabled WPCN with Separated ET and AP}
Instead of using one single UAV as co-located ET and AP, we can alternatively dispatch two different UAVs as separated ET and AP for WPT and WIT, respectively\cite{park2019uav}. The dual-UAV-enabled WPCN is expected to provide more degrees of freedom in optimizing the UAVs' trajectories to enhance the system performance. For instance, under the TDD/TDMA protocol, the UAV-ET can follow the flight trajectory in Section \ref{sec:WPT} to fairly charge the distributed GDs, and the UAV-AP can follow a different trajectory to sequentially hover above different GDs to maximally collect the information, provided that the collision avoidance issue is properly addressed. Besides, the dual-UAV-enabled WPCN also provides opportunities to enabled in-band full duplex operation between WIT and WPT, as the two UAVs can stay further away from each other during the flight to reduce the interference from the WPT of UAV-ET to the information reception at UAV-AP.

\subsubsection{UAV-Enabled Wireless Powered Backscatter Communications}
Wireless powered backscatter communications have recently attracted a lot of attention as a new type of WPCN\cite{ckhobackscatter}. Instead of using active RF chains to send information, the backscatter devices can reflect the carrier signals from the ET with properly adjusted phase and/or amplitude to convey information. In UAV-enabled wireless powered backscatter communications, the UAV can be dispatched as both ET and RF readers to not only wirelessly charge these GDs but also collect the reflected signals for information decoding at the same time. Due to the interference caused by WPT, the UAV's trajectory should be designed by considering the performance tradeoff between harvested energy and backscatter communication rate \cite{yang2019energy,Gyang2019energy,zhu2018inference,farajzadeh2020mobility,hua2019throughput,farajzadeh2019uav,yeh2020uav,hua2019uav}.


\subsubsection{Multi-UAV-Enabled WPCN}
Similarly as in multi-UAV-enabled WPT, when the network size becomes large, it becomes necessary to use multiple UAVs to implement the WPCN, in which multiple UAVs need to cooperate in the joint trajectories design, as well as the energy transmission and information reception. For instance, if the UAV swarming (cf. Section \ref{swarm}) is adopted, UAVs can cooperatively design their transmit energy covariance matrices for downlink WPT, and also use joint signal detection (via CoMP reception) for uplink WIT. On the other hand, with GD clustering, UAVs can each cover a dedicated non-overlapping sub-area for serving the GDs therein, while interference coordination needs to be considered to mitigate the co-channel interference among different clusters in WIT. Under these different designs, the UAVs' trajectories need to be properly designed jointly with the corresponding resource allocation methods. In the literature, there is one prior work \cite{xie2020common} that investigated the joint UAV trajectory design and communication resource allocation in a simplified case with two UAVs serving two GDs, where the two UAVs can cooperate in two different modes with CoMP and interference coordination, respectively. It is found in \cite{xie2020common} that in the CoMP mode, the two UAVs prefer to stay between the two GDs to enhance the cooperative beamforming gain, while in the interference coordination mode, the two UAVs would keep far away from each other to alleviate the co-channel interference in uplink WIT. How to extend the UAV-enabled WPCN to scenarios with more UAVs and more GDs under different cooperation strategies is an open problem that has not been addressed yet.




\section{UAV-Enabled Wireless Powered MEC}\label{SEC_WPMEC}
\begin{figure}
  \centering
  \includegraphics[width=10cm]{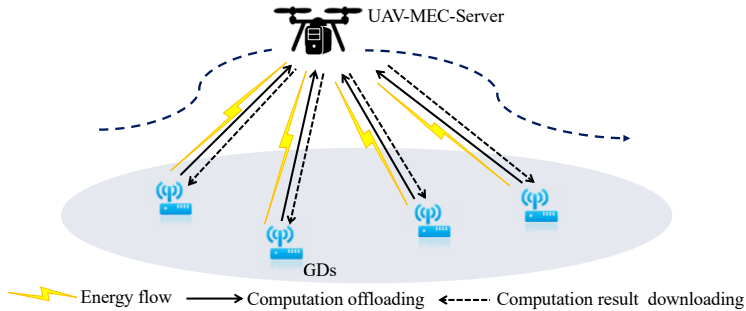}\\
  \caption{Illustration of the UAV-enabled wireless powered MEC.}\label{fig_UAV_MEC}
\end{figure}

Besides UAV-enabled WPCN, UAV-enabled wireless powered MEC is another recent application of UAV-enabled WPT, in which UAVs are dispatched as aerial MEC servers that can provide both wireless energy supply and cloud-like computing for low-power GDs.
In the single-UAV-enabled wireless powered MEC as shown in Fig. \ref{fig_UAV_MEC}, the UAV broadcasts wireless signals to charge GDs, and the GDs use the harvested energy to accomplish their respective computation tasks via local and/or remote execution.
Generally speaking, the UAV-enabled wireless powered MEC is more complicated than the UAV-enabled WPT/WPCN, as it involves WPT, WIT (for computation task offloading/downloading), and computation in a unified design.
In this case, how to design the UAV trajectory jointly with resource allocations for energy transmission, communication, and computation is a critical but challenging task. In the literature, although there have been several initial works  \cite{du2019TVT,Wang2020TNSE,liu2019IoT,zhou2018JSAC,Hu2020TWC} that investigated the UAV-enabled wireless powered MEC, the research on this topic is still in its infancy stage.

In this section, we consider the single-UAV-enabled wireless powered MEC over a particular mission period ${\mathcal T} = (0, T]$, and present a generic utility maximization problem under new computation causality constraints and energy harvesting constraints. Next, we discuss their solutions. In order to gain insights to motivate future research, we focus on the case when the computation tasks are  completely partitionable, such that the computation tasks can be partitioned into independent parts that can be executed locally (at the GD) or remotely (at the UAV) at the same time.

\subsection{Operation Protocol}



The UAV-enabled wireless powered MEC generally consists of three wireless links, including the downlink WPT from the UAV to the GDs, the uplink task offloading from the GDs to the UAV, and the downlink result downloading from the UAV to the GDs. Besides, the GDs and the UAV should implement the local and remote execution, respectively, which are causally constrained by the task offloading and result downloading. In this case, how to design an efficient operation protocol for wireless powering, communication, and computation is a complicated task.

For illustration, we consider the TDD/TDMA protocol similarly as for the UAV-enabled WPCN, in which the wireless links for downlink WPT, uplink task offloading, and downlink result downloading are implemented over the same frequency band but orthogonal time instants.
At time instant $t\in\mathcal T$, let  $\tau_E(t)\in\{0,1\}$, $\tau_k^{\rm o}(t)\in\{0,1\}$, and $\tau_k^{\rm d}(t)\in\{0,1\}$ denote the operation mode indicators. Here,  $\tau_E(t)=1$ indicates the downlink WPT mode, $\tau_k^{\rm o}(t)=1$ means that GD $k$ offloads the computation tasks to the UAV, and $\tau_k^{\rm d}(t)=1$ represents that GD $k$ downloads the computation results from the UAV.
Due to the TDD/TDMA consideration, we have $\tau_E(t)+\sum\limits_{k\in\mathcal K}\tau_k^{\rm o}(t)+\sum\limits_{k\in\mathcal K}\tau_k^{\rm d}(t)\le 1,\forall t\in\cal T$.


First, we consider the downlink WPT. As the WPT is only implemented when $\tau_E(t)=1$,
the total harvested energy at GD $k$ is given as $\bar{E}_k^{\text{tot}}(\{\mv{u}(t),\tau_E(t)\})$ in \eqref{WPCN_RP}, similarly as the UAV-enabled WPCN in Section \ref{SEC_WPCN}.


Then, we consider the task offloading form GDs to the UAV. Supposing that $Q_k(t)$ denotes the transmit power at GD $k$,  the number of task-input bits offloaded from GD $k$ to the UAV at time $t$ (in bits-per-second) is ${\ell}^{\rm off}_k(\tau_k^{\rm o}(t), Q_k(t), \mv{u}(t)) =  \tau_k^{\rm o}(t)\log_2\left(1+\frac{Q_k(t)h_k(\mv{u}(t))}{\sigma^2}\right).$

Next, after receiving the offloaded tasks from GDs, the UAV needs to execute the tasks using its computation resource and then sends the computation results back to GD $k$. In order to accomplish the offloaded tasks, the UAV must have enough computation capabilities to finish these tasks and send the computation results back to GD $k$, thus leading to the following two types of computation causality constraints. Let $f_0(t)$ denote the CPU frequency at the UAV at time $t \in \mathcal T$, $c_0$ denote the number of CPU cycles for computing one bit at the UAV. Accordingly, the computation capacity (in bits-per-second) at the UAV at each time instant $t$ is $f_0(t)/c_0$. Furthermore, let $\varphi_k(t)$ denote the computation rate (in bits-per-second) for GD $k$ at the UAV. We thus have
\begin{align}\label{MEC_bits}
	\sum\limits_{k\in\mathcal{K}} \varphi_k(t)\le \frac{f_0(t)}{c_0}, \forall t\in\mathcal T.
\end{align}
For the remote execution at the UAV, the accumulatively computed bits from for each GD $k\in\mathcal K$ by the UAV till any time instant $t$ should not exceed that accumulatively offloaded from that GD by that time. We thus obtain the first type of computation causality constraints as
\begin{align}
\int_0^t {\ell}^{\rm off}_k(\tau_k^{\rm o}(\tilde{t}), Q_k(\tilde{t}), \mv{u}(\tilde{t}))  \text{d}{\tilde t} &\ge \int_{0}^t \varphi_k(\tilde{t})\text{d}\tilde{t}, ~\forall k\in{\mathcal K},~\forall t\in\mathcal T.\label{MEC_UAV_off1}
\end{align}
Furthermore, we have the following constraint in order for the offloaded bits to be successfully computed before the deadline of $T$. 
\begin{align}\int_0^T {\ell}^{\rm off}_k(\tau_k^{\rm o}(\tilde{t}), Q_k(\tilde{t}), \mv{u}(\tilde{t})) \text{d}{\tilde t} &= \int_{0}^T \varphi_k(\tilde t)\text{d}{\tilde t}, ~\forall k\in{\mathcal K}.\label{MEC_UAV_off2}
\end{align}
Besides, after completing the remote task execution, the UAV needs to download the computation results back to GDs.
Under transmit power $P$ at the UAV, the number of downloaded bits at time instant $t$ is
${\ell}^{\rm down}_k(\tau_k^{\rm d}(t), \mv{u}(t)) = \tau_k^{\rm d}(t)\log\left(1+\frac{Ph_k(\mv{u}(t))}{\sigma_k^2}\right),$ where $\sigma_k^2$ denotes the noise power at the receiver of GD $k$. Notice that the accumulatively downloaded bits should not exceed that accumulatively computed at each time $t\in\mathcal T$ and the downloading must be accomplished before the deadline of $T$. Therefore, by supposing that the size of computation results (task output bits) is a $\beta$ portion of task input bits, we further have the second type of computation causality constraints as
\begin{align}
\int_0^t {\ell}^{\rm down}_k(\tau_k^{\rm d}(\tilde{t}), Q_k(\tilde{t}), \mv{u}(\tilde{t}))  \text{d}{\tilde t} &\le \beta\int_{0}^t \varphi_k(\tilde{t})\text{d}\tilde{t}, ~\forall k\in{\mathcal K},~\forall t\in\mathcal T,\label{MEC_UAV_down1}\\
\int_0^T {\ell}^{\rm down}_k(\tau_k^{\rm d}(\tilde{t}), Q_k(\tilde{t}), \mv{u}(\tilde{t})) \text{d}{\tilde t} &=\beta \int_{0}^T \varphi_k(\tilde t)\text{d}{\tilde t}, ~\forall k\in{\mathcal K}.\label{MEC_UAV_down2}
\end{align}


Finally, we consider the local computing at GDs. Let $f_k(t)$ denote the CPU frequency (in cycles per second) at time $t\in\mathcal T$, and $c_k$ the number of CPU cycles for computing one bit at GD $k$.
Accordingly, the computation rate at GD $k$ (in bits per second) is $ f_k(t)/c_k$, and the corresponding power consumption is $ \kappa_k f_k^3(t)$ \cite{Cao2019IoT}, where $\kappa_k$ deno     tes the effective capacitance coefficient depending on the chip architecture. Hence, the total number of computation bits executed at GD $k$ locally is $L^{\rm loc}_k(\{f_k(t)\})= \int_0^T f_k(t)/c_k \text{d}t $.

\subsection{Joint Trajectory and Resource Allocation Design for Computation Utility Maximization}

Under the TDD/TDMA protocol, we are interested in maximizing the computation rate or the total number of computed bits at each GD, given by  $ L^{\rm off}_k(\{\tau_k^{\rm o}(t), Q_k(t), \mv{u}(t)\})+L^{\rm loc}_k(\{f_k(t)\})$, where $L^{\rm off}_k(\{\tau_k^{\rm o}(t), Q_k(t), \mv{u}(t)\})=\int_0^T {\ell}^{\rm off}_k(\{\tau_k^{\rm o}(t), Q_k(t), \mv{u}(t)\})\text{d}t$.
 Accordingly, we define the utility function as
\begin{align}
{\tilde U}(\{\tau_k^{\rm o}(t), Q_k(t), \mv{u}(t),f_k(t)\})=\min_{k\in\mathcal K}\left\{\frac{L^{\rm off}_k(\{\tau_k^{\rm o}(t), Q_k(t), \mv{u}(t)\})+L^{\rm loc}_k(\{f_k(t)\})}{\tilde a_k}\right\},
\end{align}
where $\tilde a_k$ denotes the computing weight for each GD $k$.

Notice that similarly as in \eqref{energycau} for UAV-enabled WPCN, each GD $k$ is subject to the energy harvesting constraints. By combining the communication energy consumption $Q_k(t)$ for offloading and computation energy consumption $\kappa_k f_k^3(t)$ for local task execution, the energy harvesting constraints can be expressed as
\begin{align}\label{MEC_Device_energy_casual}
\int_0^t \kappa_k f_k^3({\tilde t})\text{d}{\tilde t} +\int_0^{t} \tau_k({\tilde t})Q_k({\tilde t})\text{d}{\tilde t}\leq \int_0^{t} \tau_E({\tilde t})\eta  P h_k(\mv{u}({\tilde t}))\text{d}{\tilde t}+ E_k^{\rm initial},\forall k\in\mathcal K, t\in\mathcal T,
\end{align}
or the following one if the initial energy storage $E_k^{\rm initial}$ becomes sufficiently large.
\begin{align}\label{MEC_Device_energy_total}
\int_0^T \kappa_k f_k^3(t)\text{d}t +\int_0^{T} \tau_k(t)Q_k(t)\text{d}t \leq \int_0^{T} \tau_E(t)\eta  P h_k(\mv{u}(t))\text{d}t,\forall k\in\mathcal K, t\in\mathcal T.
\end{align}

In this case, we have the utility maximization problem as
\begin{align} \text{(P4):}~\max_{\{\mv{u}(t),Q_k(t)\ge0,f_0(t)\ge0,f_k(t)\ge 0,\varphi_k(t) \ge0 \}}~&{\tilde U}(\{\tau_k^{\rm o}(t), Q_k(t), \mv{u}(t),f_k(t)\})\notag\\
	\mathtt{s.t.}~~~~~~~~~~~~~~~~~~&\eqref{P1con},~\eqref{MEC_bits},~ \eqref{MEC_UAV_off1},~\eqref{MEC_UAV_off2},~\eqref{MEC_UAV_down1},~\eqref{MEC_UAV_down2},~\text{and}~\eqref{MEC_Device_energy_casual}.\notag
\end{align}

Notice that problem (P4) for joint UAV trajectory and resource allocation design in the UAV-enabled wireless powered MEC is more difficult to solve than problem (P3) in the UAV-enabled WPCN, due to the computation causality constraints. How to solve problem (P4) is a challenging problem that has not been addressed yet.

In the literature, there have been some initial works \cite{du2019TVT,liu2019IoT,zhou2018JSAC,Hu2020TWC} that investigated simplified versions of problem (P4) by ignoring the computation causality constraints in \eqref{MEC_UAV_off1}--\eqref{MEC_UAV_down2}. Indeed, in this case, problem (P4) has a similar structure as problem (P3), with the newly considered computation resource allocation variables.
As a result, we can use similar approaches (i.e., multi-location-hovering, SHF trajectory, and time-quantization-based optimization) as those for problem (P3) in Section \ref{Pro3_solu} to solve the simplified version of (P4). For the general problem (P4) with constraints \eqref{MEC_UAV_off1}--\eqref{MEC_UAV_down2} considered, however, the above approaches may not work well due to the computation causality constraints.
To tackle this issue, we may directly use time-quantization to transform problem (P4) with continuous-time variables as equivalent problems in discrete time, and then use the SCA techniques to solve it. It is expected that the UAV may fly back and forth to visit different GDs for offloading and downloading over time due to the computation causality constraints (see, e.g., \cite{XCao18UAVMEC}).

\subsection{Extensions}


The single-UAV-enabled wireless powered MEC design can also be extended to the case with multiple UAVs and cloud integration, as discussed in the following to motivate future research.

\begin{itemize}
	\item {\it Multi-UAV Coordination:} To provide sustainable computation services in a large area, multiple UAVs can be dispatched to cooperatively serve a large number of low-power GDs. In this case, how to associate GDs with these UAVs and properly schedule their energy transmission, computation, and communication resources is a new problem to tackle.
	\item {\it Edge-Cloud Integration:} As the UAV-MEC-servers generally have limited computation capabilities, it is desirable to further integrate the centralized clouds to help the task execution, especially when the computation tasks are heavy.  In practice, the centralized cloud can either be a large data center on the ground or deployed in high-altitude platforms or even satellites \cite{WChen19Network}.
	In this case, it is necessary to partition the computation tasks into different parts to be executed locally at GDs and remotely at the UAV-MEC-server and/or cloud.
\end{itemize}

\section{Other Extensions}\label{SEC_Extension}

In the preceding sections, we provided an overview on the UAV-enabled WPT and its applications in UAV-enabled WPCN and UAV-enabled wireless powered MEC. Due to the space limitation, there are several important issues that are unaddressed. In the following, we briefly discuss these issues to motivate future research.

\subsection{Non-linear Energy Harvesting Model}
So far, we focused on the approximate linear energy harvesting model at GDs as widely adopted in the literature. In practice, however, the harvested DC power may not be a linear function with respect to the received RF power, especially when the received RF power level is sufficiently high or low.
In general, the effect of non-linear energy harvesting models on the UAV-enabled WPT is still an uncharted area in the literature.
For instance, due to the energy saturation at high RF power, the UAV may not need to fly exactly above the GDs for most efficient charging.
 Therefore, how to jointly optimize the transmit waveform and UAV trajectory is a crucial issue to be tackled for UAV-enabled WPT by considering the non-linear energy harvesting model\cite{UAVnonlinear}.

\subsection{CSI Availability}
CSI is important for UAVs to implement the (cooperative) transmit energy beamforming for WPT in multi-antenna and multi-UAV scenarios. There are generally three approaches for obtaining CSI in WPT, namely  energy feedback, reverse-link channel estimation based on pilots, and channel estimation with limited feedback \cite{Xu14TSP_onebit}. To obtain the CSI, the GDs normally need to consume additional time and energy to send training signals or implement channel estimation/feedback.
Therefore, there generally exists a fundamental tradeoff in obtaining accurate CSI for efficient energy beamforming versus minimizing time/energy consumption\cite{optimizednetZeng,optimizedZeng}.

\subsection{AirComp}
AirComp is an emerging approach to enable fast wireless data aggregation for achieving functional computation over the air by exploiting the superposition property of multiple access channels \cite{nomo_function_Nazer,Cao_PowerTWC,Cao_2020aa}.  By exploiting AirComp, UAV and WPT, the UAV-enabled wireless powered AirComp can be an efficient way for aggregating the data from low-power GDs, which is expected to have abundant applications in future massive machine type communications, and for distributed learning \cite{GZhu2020TWC}, sensing \cite{Abari2016Ar_AirComp}, and consensus \cite{FMolinari18_Consensus}.
The key challenge lies in how to fully exploit the mobility of UAVs to achieve the required signal alignment for AirComp in a sustainable operation manner.

\subsection{Online Trajectory Design}
So far, the UAV trajectory design is formulated as deterministic optimization problems, which can be generally solved in an offline manner via multi-location hovering, SHF trajectory design, time quantization and their variants. These designs, however, cannot work well when the channel propagation environments are spatial- and time-varying due to the obstacles between UAVs and moving GDs.
To tackle this challenge, radio map \cite{Zhang20SPWAC_online,Esrafilian19IoT_map,XMo20WCL_WPT} and reinforcement learning (RL) \cite{Bayerlein20GC_online} are two useful tools to help autonomously update the UAV trajectories in an online manner.


\subsection{Ground Vehicles for WPT}
Besides UAVs, ground vehicles can also be dispatched to serve as another type of mobile ETs on the ground to facilitate WPT, and the trajectory design framework can be generally extended to design the moving trajectory of ground vehicles for efficient WPT.
Nevertheless, unlike UAVs that can freely fly in the 3D airspace, ground vehicles need to travel in the prescribed lane according to road conditions. Therefore, the trajectory optimization of ground vehicles is less flexible and may be even more challenging than that of UAVs.

%

\section{Conclusion}\label{SEC_Con}

In this paper, we provided a tutorial overview on UAV-enabled WPT as well as its various applications and extensions, by focusing on how to exploit the UAV mobility to enhance the system performance.
First, in the single-UAV-enabled WPT case, we presented a trajectory design framework to fairly maximize the harvested energy at multiple GDs, which consists of three main approaches, namely multi-location hovering, successive-hover-and-fly, and time-quantization-based optimization.
Next, we extended the trajectory design framework to the multi-UAV-enabled WPT case based on the schemes of UAV swarming and GD clustering, respectively.
Then, we considered the UAV-enabled WPCN and wireless powered MEC, in which the trajectory design framework is investigated jointly with the resource allocations to improve communication/computation performance.
Furthermore, open problems and promising research directions in UAV-enabled WPT were presented to inspire future exploration.

\bibliography{myref,myref_wpcn}
\bibliographystyle{IEEEtran}
\end{document}